\documentclass[english]{article}
\usepackage[T1]{fontenc}
\usepackage[latin9]{inputenc}
\usepackage{geometry}
\usepackage{amsmath}
\usepackage{amssymb}
\usepackage{stackrel}
\usepackage{graphicx}

\makeatletter

\providecommand{\tabularnewline}{\\}

\newcommand{\lyxaddress}[1]{
	\par {\raggedright #1
	\vspace{1.4em}
	\noindent\par}
}

\date{}

\makeatother

\usepackage{babel}
\begin{document}
\title{Transport in Hamiltonian systems with slowly changing phase space
structure}
\author{Freddy Bouchet$^{1}$ and Eric Woillez$^{1,2}$\footnote{Corresponding author. \emph{Email address}: eric.woillez@technion.ac.il} }
\maketitle

\lyxaddress{1) Univ Lyon, Ens de Lyon, Univ Claude Bernard, CNRS, Laboratoire
de Physique, F-69342 Lyon, France.\\
2) Department of Physics, Technion, Haifa 32000, Israel.}
\begin{abstract}
Transport in Hamiltonian systems with weak chaotic perturbations has
been much studied in the past. In this paper, we introduce a new class
of problems: transport in Hamiltonian systems with slowly changing
phase space structure that are not order one perturbations of a given
Hamiltonian. This class of problems is very important for many applications,
for instance in celestial mechanics. As an example, we study a class
of one-dimensional Hamiltonians that depend explicitly on time and
on stochastic external parameters. The variations of the external
parameters are responsible for a distortion of the phase space structures:
chaotic, weakly chaotic and regular sets change with time. We show
that theoretical predictions of transport rates can be made in the
limit where the variations of the stochastic parameters are very slow
compared to the Hamiltonian dynamics. Exact asymptotic results can
be obtained in the classical case where the Hamiltonian dynamics is
integrable for fixed values of the parameters. For the more interesting
chaotic Hamiltonian dynamics case, we show that two mechanisms contribute
to the transport. For some range of the parameter variations, one
mechanism -called "transport by migration with
the mixing regions" - is dominant. We are then able to
model transport in phase space by a Markov model, the local diffusion
model, and to give reasonably good transport estimates. 
\end{abstract}

\section{Introduction}

Transport in Hamiltonian systems and is a classical field of dynamical
system theory \cite{zaslavsky2002chaos,lichtenberg2013regular,wiggins2013chaotic,meiss2015thirty},
with a huge number of applications \cite{elskens2002microscopic,escande2018thermonuclear}.
Beyond deterministic dynamical systems, a lot of work has been devoted
in the past to study the effect of random perturbations \cite{moss1989noise},
more specifically on Hamiltonian systems and on area preserving maps,
for instance in the context of plasma physics \cite{rechester1980calculation,rechester1981fourier}.
Noise is always present in real natural systems and in experiments
because of the effect of hidden chaotic degrees of freedom. Even of
small amplitude, noise plays a very important role in the long term
behavior of the dynamics, and on transport properties. One usually
models the hidden degrees of freedom by an additional stochastic process
of small amplitude acting on the system. For example, the effect of
noise on the standard map or on other classical area preserving maps
has been studied earlier in \cite{karney1982effect,lieberman1972stochastic,lichtenberg2013regular},
motivated by the dynamics of charged particles in accelerators. More
recently, with the development of stochastic calculus, \cite{freidlin1984random,freidlin2008some}
have studied the generic effect of small stochastic perturbations
of Hamiltonian flows, and \cite{bazzani1997action,bazzani1998diffusion}
have derived a diffusion equation for the slow action variable in
Hamiltonian systems. In particular, \cite{freidlin1984random,freidlin2008some}
have rigorously justified the averaging principle in Hamiltonian systems
and studied the slow diffusive motion of action variables. The important
point is that all those works fall in the dynamical framework
\begin{equation}
\dot{x}=\frac{1}{\epsilon}F(x)+\beta\left(x,\frac{t}{\epsilon}\right),\label{eq:exemple generic}
\end{equation}
where $F$ is a Hamiltonian vector field, $\beta$ is a deterministic
or a stochastic perturbation of the Hamiltonian vector field, and
$\epsilon$ is a small parameter. Qualitatively, we can describe the
dynamics of (\ref{eq:exemple generic}) saying that it follows the
regular or chaotic orbits of the unperturbed dynamics $\dot{x}=F(x)$ on a fast timescale $\propto \epsilon$,
and deviates slowly from those orbits because of the effect of the
perturbation $\beta$, acting on a timescale of order one.  We note that the case  (\ref{eq:exemple generic}) where $F$ is a Hamiltonian dynamics and $\beta$ is a wave of slow modulated frequency has been studied in \cite{PhysRevE.75.065201,Makarov2010,Uleysky2010,makarov2012control} and can lead to interesting phenomena such as autoresonant motion and acceleration of particles.

In this paper, we consider a different framework, which cannot be reduced
to the much-studied model (\ref{eq:exemple generic}), and still is
essential for many applications. We study a one-dimensional Hamiltonian
dynamics which Hamiltonian depends on a slow parameter\footnote{Note that we could have defined a new Hamiltonian $H'=H/\epsilon$ such that Eq. (\ref{eq:hamiltonien modele}) are indeed Hamilton's equations with $H'$, but we believe the presentation chosen in this paper emphasizes more clearly the timescale separation in the dynamics.}
\begin{equation}
\dot{x}=\frac{1}{\epsilon}J\nabla H(x,\nu(t)),\label{eq:hamiltonien modele}
\end{equation}
where $x:=(p,q)$ represents the vector of canonical variables, and
$J:=\begin{bmatrix}0 & -1\\
1 & 0
\end{bmatrix}$. 

\begin{figure}
\includegraphics[height=4.3cm]{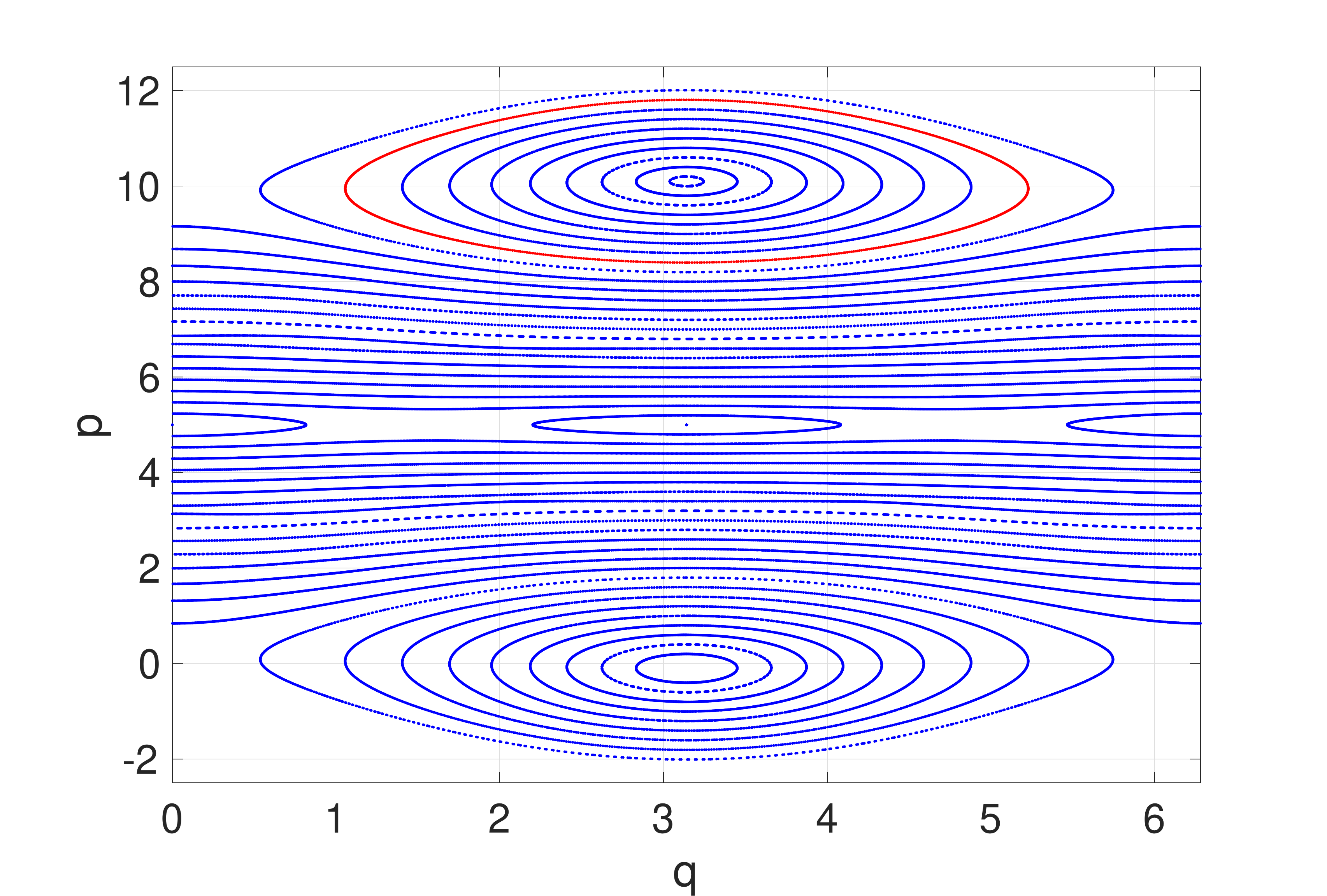}\includegraphics[height=4.3cm]{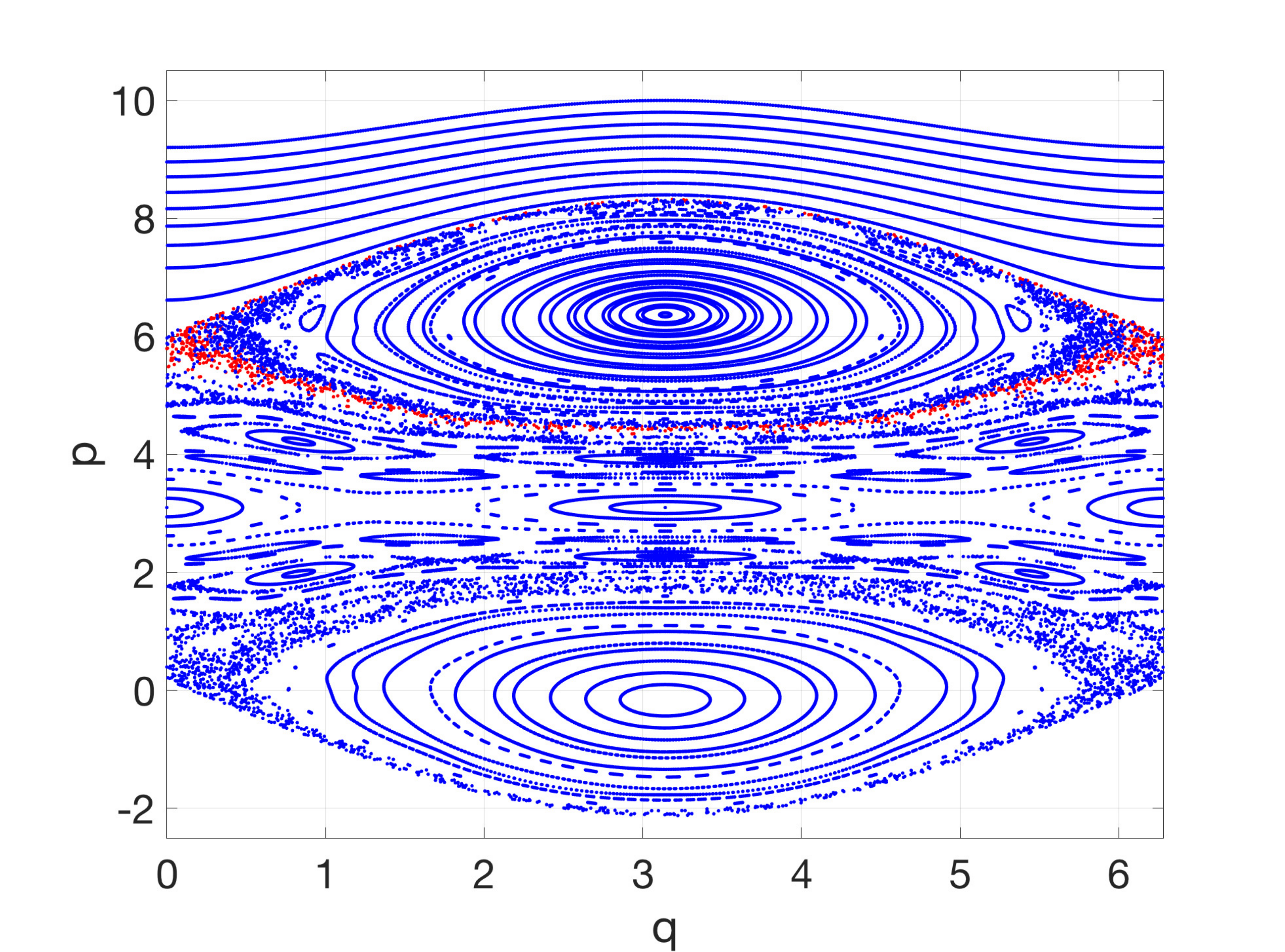}\includegraphics[height=4.3cm]{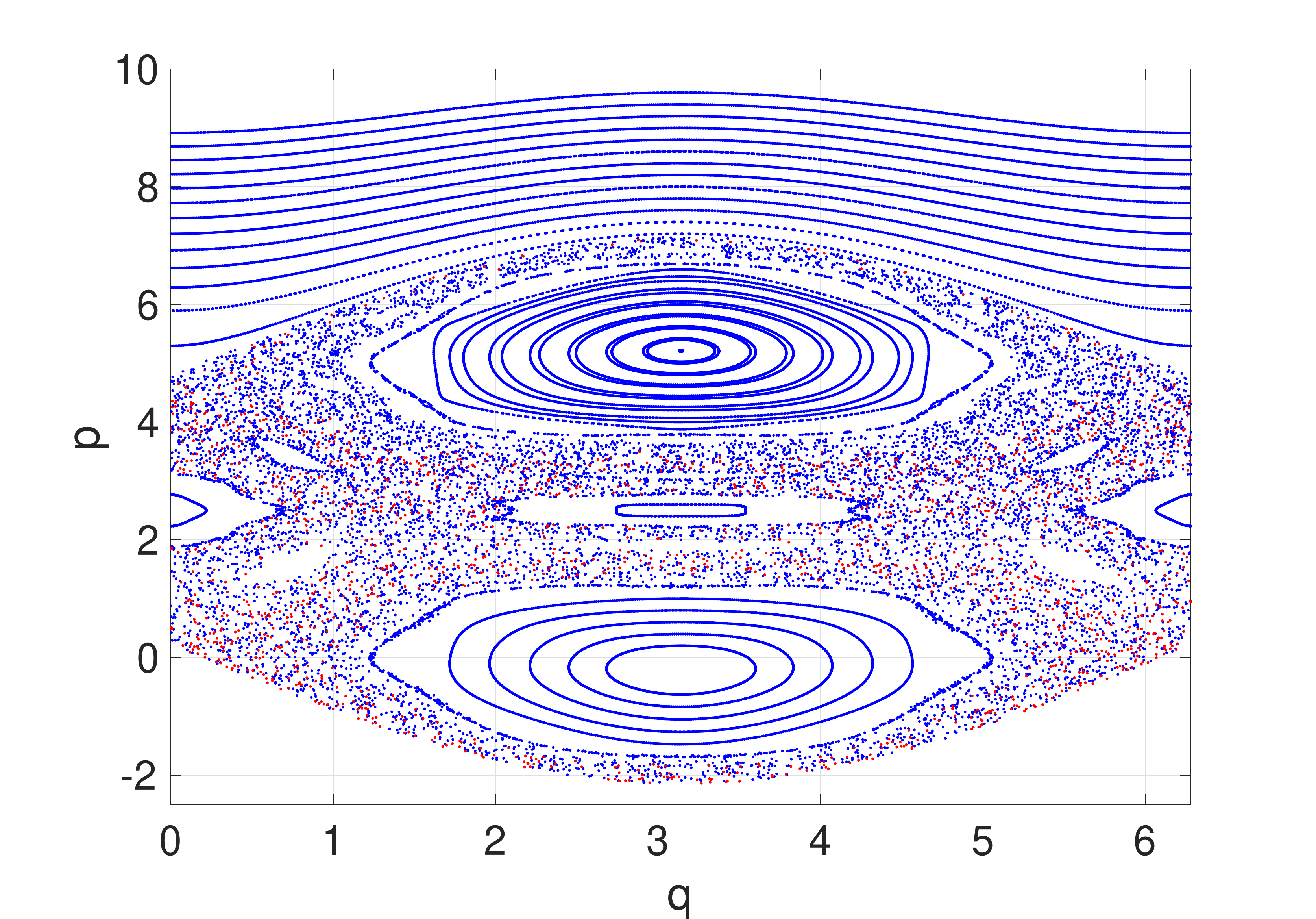}

\caption{Snapshots of the phase space structure of the Hamiltonian (\ref{eq:Hamiltonien demo})
for three different values of $\nu$. The blue curves represent a
Poincar\'e section of the Hamiltonian flow (\ref{eq:Hamiltonien demo}).
The position of the lower resonance is fixed, but the upper resonance
is centered at the value $p=\nu$. Left panel: $\nu\approx10$, the
two main resonances are far away from each other and the phase space
is very weakly chaotic. The red trajectory is trapped in the upper
libration region. Middle panel: $\nu\approx6.2$, and chaotic regions
become larger in the vicinity of hyperbolic fixed points. The red
trajectory is in a chaotic region, but it cannot be carried to the
lower libration region because of remaining KAM tori. Right panel:
$\nu\approx5$ and the two main cat's eyes are now connected through
a strongly chaotic region. KAM tori have been destroyed and the red
trajectory can freely do the transition from one libration region
to the other. \label{fig:demo}}
\end{figure}

This model has received relatively little attention in the literature compared to Eq. (\ref{eq:exemple generic}), but some nontrivial features have already been emphasized in \cite{brannstrom2008drift,gelfreich2008unbounded} such as the existence of trajectories with unbounded energy growth. Generally the dynamics (\ref{eq:exemple generic}) can be very different
from the dynamics (\ref{eq:hamiltonien modele}) because the variations
of $\nu$ in Hamiltonian (\ref{eq:hamiltonien modele}) can range over a region of order one. The amplitude of the variations of $\nu$ can thus be of the same order as the variations of action variables. We call $\nu$ a "slow variable", because one has to wait for a time $\Delta t$ of order one to observe a variation $\Delta \nu$ of order one, whereas canonical variables have large variations on a timescale $\propto \epsilon$. For any fixed value of the parameter
$\nu$, for the Hamiltonian $H\left(x,\nu\right)$, the dynamics
in its phase space is characterized by strongly chaotic regions, weakly
chaotic regions, and in some cases, KAM tori. We call phase space
structure the geometry and topology of these chaotic, weakly chaotic,
and regular areas. When the parameter $\nu$ slowly evolves with time,
the geometry and topology of those regions are slowly distorted, and
can be dramatically changed for large variations of $\nu$ that occur
on long times. As a consequence of the distortion of the phase space
structure, some regions that might not have been accessible for the
system at the initial value of $\nu$ become easily accessible when
$\nu$ changes. This affects drastically the transport in phase space.
Let us consider a simple example to illustrate such a change in phase
space structure. We take the one and a half degree of freedom Hamiltonian\footnote{Note that we call this model one and a half degree of freedom Hamiltonian because $\Lambda$ do not play any role in the dynamics.} 
\begin{equation}
H(p,q,\Lambda, \lambda, \nu)=\frac{p^{2}}{2}+\cos\left(q\right)+\cos\left(q-\lambda\right) +\nu\Lambda,\label{eq:Hamiltonien demo}
\end{equation}
where $\nu$ is the frequency of the angle $\lambda$ and plays the role of the external parameter and is the conjugated angle to $\Lambda$. 
In the Hamiltonian (\ref{eq:Hamiltonien demo}), a resonance is defined
as the value of $p$ for which one of the two angles $q$ or $q+\lambda$
has zero frequency. We have plotted in figure (\ref{fig:demo})
a snapshot of the phase space structure $(p,q)$ for the Hamiltonian (\ref{eq:Hamiltonien demo})
for three different values of $\nu$. The frequency $\nu$ is decreasing
from the left picture to the right picture. The Poincar\'e section described
by (\ref{eq:Hamiltonien demo}) displays two major libration regions,
one centered at $p=0$, and one centered at $p=\nu$. A libration
region is defined as a region of phase space where the canonical angle
$q$ has bounded oscillations between two extremal values in $[0,2\pi]$.
The libration regions can be very easily recognized on the Poincar\'e
section of figure (\ref{fig:demo}) as they look like \textquotedbl cat's
eyes\textquotedbl{} surrounding the fixed points of resonances. The
Hamiltonian flow is only very weakly chaotic in the initial state
because the two main resonances are far away from each other, but
it becomes more and more chaotic as the two main resonances become
closer (for a precise description of such systems, see \cite{lichtenberg2013regular}).
The red curves in figure (\ref{fig:demo}) are particular trajectories
that start inside the upper libration region. For large enough values
of $\nu$, the trajectories are trapped inside the upper libration
region (first and second panel of figure (\ref{fig:demo})). On the
contrary, when $\nu$ is lower than some critical value, the trajectory
can freely transit from one cat's eye to the other (third panel of
figure (\ref{fig:demo})). Imagine now that the frequency $\nu$ in
Hamiltonian (\ref{eq:Hamiltonien demo}) is no longer fixed, but slowly
depends on time. Consider then the Hamiltonian
\begin{equation}
H(p,q,\Lambda,\lambda,\nu(t))=\frac{p^{2}}{2}+\cos\left(q\right)+\cos\left(q-\lambda\right) +\nu(t)\Lambda.\label{eq:Hamiltonien demo-1}
\end{equation}

The three pictures of figure (\ref{fig:demo}) represent instantaneous
snapshots of the phase space of the Hamiltonian (\ref{eq:Hamiltonien demo-1})
at three different times. The upper cat's eye centered at $p=\nu$
moves according to the variations of $\nu(t)$. Some trajectories
that were trapped in the upper libration region in the initial state
can be carried downward by the displacement of the upper libration
region and finally reach the lower libration region. \\

The aim of the example of figure (\ref{fig:demo}) is to show that
the slow changes in phase space structure create a particular transport
mechanism that is very different from the transport created by small
perturbation as in system (\ref{eq:exemple generic}). In the present
paper, we investigate the case (\ref{eq:hamiltonien modele}) where
the parameter $\nu$ is \emph{stochastic}. In particular, we consider in the present work the possibility of strongly irregular functions $\nu(t)$, as happens for example when $\nu$ is a diffusion process. This irregularity can be seen, for example, as the result of the influence of chaotic external degrees of freedom that evolve on a timescale much shorter than any other timescale in the Hamiltonian system.
We show that transport
in phase space is the result of two mechanisms. We call the first
mechanism \emph{noise driven transport in regular regions.} This transport
mechanism comes from the strong irregularity of the stochastic trajectories
$\nu(t)$, and would not be observed for infinitely differentiable functions $\nu(t)$. We call the second mechanism \emph{transport by slow deformation
of chaotic regions.} This second transport mechanism occurs when trajectories
are trapped in chaotic regions and follow their displacement to access
to other parts of phase space. During the displacement, the extension of the chaotic layer can also vary such that transport is due to both \emph{migration} and \emph{extension} of chaotic regions,  as happens for example in Fig. (\ref{fig:demo}). For the generic dynamics (\ref{eq:hamiltonien modele})
both mechanisms, noise driven transport in regular regions and transport by slow deformation
of chaotic regions, are present and contribute to transport. Our aim is
to separately describe and quantify transport due to the one and the
other mechanisms. 

All over the paper, we thus consider the Hamilton's equations (\ref{eq:hamiltonien modele}).
We first consider in section
\ref{sec:The-integrable-case} the case of a one-degree of freedom
Hamiltonian of the form $H(p,q,\nu(t)),$ which has an integrable
dynamics for any fixed value of $\nu$. If the parameter $\nu$
 would be a regular function of time, transport would be described by the
classical theory of adiabatic invariants \cite{kulsrud1957adiabatic,gardner1959adiabatic}.
For a \emph{stochastic} parameter $\nu$ with irregular trajectories, we show that the transport
is completely due to the mechanism of noise driven transport in regular
regions, and we give a stochastic differential equation to describe
the diffusion of adiabatic invariants in the limit $\epsilon\rightarrow0$, when there is no separatrix crossing during the diffusion process. 
This result is obtained using standard averaging techniques. We also
illustrate our results with the example of the harmonic
oscillator. We note that the case of diffusion across a separatrix is much more technical. It has been shown in case of a dynamics of type (\ref{eq:exemple generic})  with separatrix crossing that the motion of the slow variable is given by a diffusion process on a graph where the branches represent the different domains in phase space defined by the separatrix \cite{freidlin2008some}. The change in the value of the adiabatic invariant during separatrix crossing has also been known for a long time \cite{neishtadt1987change,tennyson1986change}, and the result could in principle be used to extend our theory also to separatrix crossing for Hamiltonians with one degree of freedom. We do not treat this problem in the present paper as we have chosen to rather deal with the chaotic case where resonances overlap.\\

The second part \ref{sec:The-chaotic-case} of the paper is probably
the most original and interesting one. We deal with the much more
difficult case where the Hamiltonian dynamics (\ref{eq:hamiltonien modele})
has more than one degree of freedom and is thus chaotic even for fixed
values of $\nu$. For simplicity, we study a dynamical system where the width of chaotic regions is almost constant and the deformation only comes from migration of the regions. Our approach could be used in the general case without further difficulties. We show that both transport mechanisms, \textquotedbl noise
driven transport in regular regions\textquotedbl{} and \textquotedbl transport
by slow migration of chaotic regions\textquotedbl , take place. We
explain that the second mechanism is dominant for some range of parameters.
The transport can then be reduced to a fully Markovian process in
the limit $\epsilon\rightarrow0$. We call this process the \emph{instantaneous
local diffusion model, }because it consists in modeling the chaotic
regions of phase space by diffusive regions with infinite diffusion
coefficient. We check numerically that the Markov model gives reasonably
good estimation of transport rates when the mechanism of transport
by slow migration of chaotic regions is dominant.

Our results are relevant in celestial mechanics to study the long
term dynamics of the spin axis of planets, and the secular dynamics
of Mercury. Some times ago \cite{henrard1982capture} already pointed
out that the dynamics of the obliquity of Mars could be reduced to
a model of a pendulum with slowly varying parameters. He used the
theory of adiabatic invariants to estimate the probability of Mars
to enter into libration. However, this model still considered that
the parameters are regular functions of time. With the work of \cite{Laskar1989,laskar1990chaotic},
it has been shown that the solar system is chaotic on a timescale
of few million years. The set of fundamental frequencies of the Laplace\textendash Lagrange
solution, that plays the role of parameters in the Hamiltonian describing
the dynamics of spin axes of planets, has thus a stochastic long-term
evolution \cite{laskar2004long}. This justifies the relevance of
a model such as (\ref{eq:hamiltonien modele}). This paper was thought
of as a theoretical work to put in a general framework the ideas developed
in \cite{woillez2018complex} to study the long term dynamics of the
Earth obliquity.

\section{The integrable case\label{sec:The-integrable-case}}

\subsection{Formulation of the model and theoretical results\label{subsec:theory}}

In this section, we consider a one-degree of freedom Hamiltonian $H(p,q,\nu)$ and Hamilton's equations (\ref{eq:hamiltonien modele}), which are explicitly 
\begin{equation}
\begin{cases}
\dot{p} & =-\frac{1}{\epsilon}\frac{\partial H}{\partial q}\left(p,q,\nu(t)\right)\\
\dot{q} & =\frac{1}{\epsilon}\frac{\partial H}{\partial p}\left(p,q,\nu(t)\right)
\end{cases}\label{eq:Hamilton}
\end{equation}
The set $(p,q)$ is the set of canonical variables, and $\epsilon\ll 1$. $\nu$ is an external parameter
in the Hamiltonian. For any fixed value of the parameter $\nu$, the
Hamiltonian has one degree of freedom, and its dynamics is thus integrable.
When the parameter $\nu$ is a regular, slowly varying function of
time, the slow action dynamics is described by the old theory of adiabatic
invariants. Even when transport occurs through separatrix crossing,
the theory of adiabatic invariants can be extended to account for
the discontinuity in the action definition (see \cite{neishtadt1987change,tennyson1986change} and \cite{bazzani2014analysis}
for a review). 

We consider here a case where the parameter $\nu$
is stochastic, and we assume that no
separatrix crossing occurs during transport. The process $\nu(t)$ is in general non-differentiable, because it displays variations on arbitrary small timescales (like e.g. the standard Brownian motion).  We call $\nu(t)$ a "slow process" in the following sense: over a time interval $\Delta t$ of order $\epsilon$, the variations of the stochastic process scale as $\Delta \nu\propto \sqrt{\epsilon\Delta t}$, and are thus very small compared to the variations of canonical variables. Variations $\Delta\nu$ of order one only occur on a timescale $\Delta t$ of order one. With this signification, $\epsilon$  creates  a timescale separation between the
Hamiltonian dynamics and the stochastic dynamics of $\nu$.  We call $\epsilon$ the "fast timescale", as opposed to the natural timescale of order one to describe the system, that we call the "slow timescale". The time evolution of $\nu(t)$ is given by
\begin{equation}
{\rm d}\nu=a(\nu){\rm d}t+b(\nu){\rm d}W.\label{eq:dynamics}
\end{equation}
In the stochastic differential equation (\ref{eq:dynamics}), the
stochastic product $b(\nu){\rm d}W$ is defined with the It\^o convention
of stochastic calculus (see \cite{gardiner1985stochastic} chapter
4). We again strongly emphasize that the variable $\nu(t)$ is not a slow variable in the usual sense, because its motion contains infinitely fast variations. It can only be considered as "slow" in the sense that its time-integrated variations $\Delta \nu=\int_0^{\Delta t} \dot{\nu}(s) ds$ are much smaller than the variations of $\{p,q\}$ over the same time interval.

As the Hamiltonian $H(p,q,\nu)$ is integrable for fixed $\nu$, we
can find a set of action-angle canonical variables $(P,Q)$. With
this change of variables, the new Hamiltonian $\widetilde{H}(P,\nu)$
does not depend on $Q$. If the value of $\nu$ is fixed, the action
variable $P(p,q,\nu)$ is a constant of motion under the dynamics
of the Hamiltonian $\widetilde{H}$. In the model defined by equations
(\ref{eq:Hamilton}-\ref{eq:dynamics}), the parameter $\nu$ follows a stochastic differential equation.
This implies that the action variable $P$ also follows a stochastic
differential equation, and evolves on the same timescale as $\nu(t)$. Using
the principles of It\^o stochastic calculus, the stochastic differential
equations on $P$ and $Q$ are
\begin{equation}
\begin{cases}
{\rm d}P & =-\frac{1}{\epsilon}\frac{\partial P}{\partial p}\frac{\partial H}{\partial q}{\rm d}t+\frac{1}{\epsilon}\frac{\partial P}{\partial q}\frac{\partial H}{\partial p}{\rm d}t+\frac{\partial P}{\partial\nu}\left[a(\nu){\rm d}t+b(\nu){\rm d}W\right]+\frac{1}{2}\frac{\partial^{2}P}{\partial\nu^{2}}b^{2}(\nu){\rm d}t,\\
{\rm d}Q & =-\frac{1}{\epsilon}\frac{\partial Q}{\partial p}\frac{\partial H}{\partial q}{\rm d}t+\frac{1}{\epsilon}\frac{\partial Q}{\partial q}\frac{\partial H}{\partial p}{\rm d}t+\frac{\partial Q}{\partial\nu}\left[a(\nu){\rm d}t+b(\nu){\rm d}W\right]+\frac{1}{2}\frac{\partial^{2}Q}{\partial\nu^{2}}b^{2}(\nu){\rm d}t.
\end{cases}\label{eq:eq1}
\end{equation}
Because the set $(P,Q)$ is a set of action-angle variables, the term
$-\frac{\partial P}{\partial p}\frac{\partial H}{\partial q}{\rm d}t+\frac{\partial P}{\partial q}\frac{\partial H}{\partial p}{\rm d}t$
vanishes. This comes as a consequence of the fact that $P$ is constant
if $\nu$ is constant. The term $-\frac{\partial Q}{\partial p}\frac{\partial H}{\partial q}{\rm d}t+\frac{\partial Q}{\partial q}\frac{\partial H}{\partial p}{\rm d}t$
defines the dynamics of the angle variable $Q$ when $\nu$ is fixed,
and thus 
\[
-\frac{\partial Q}{\partial p}\frac{\partial H}{\partial q}{\rm d}t+\frac{\partial Q}{\partial q}\frac{\partial H}{\partial p}{\rm d}t=\omega(P,\nu){\rm d}t,
\]
where we have defined the pulsation $\omega(P,\nu):=\frac{\partial\widetilde{H}}{\partial P}(P,\nu)$.
Equations (\ref{eq:dynamics}-\ref{eq:eq1}) become 
\begin{equation}
\begin{cases}
{\rm d}P & =\left[\frac{\partial P}{\partial\nu}a(\nu)+\frac{1}{2}\frac{\partial^{2}P}{\partial\nu^{2}}b^{2}(\nu)\right]{\rm d}t+\frac{\partial P}{\partial\nu}b(\nu){\rm d}W,\\
{\rm d}Q & =\frac{1}{\epsilon}\omega(P,\nu){\rm d}t+\left[\frac{\partial Q}{\partial\nu}a(\nu)+\frac{1}{2}\frac{\partial^{2}Q}{\partial\nu^{2}}\epsilon b^{2}(\nu)\right]{\rm d}t+\frac{\partial Q}{\partial\nu}b(\nu){\rm d}W,
\end{cases}\label{eq:system slow-fast}
\end{equation}
and
\[
{\rm d}\nu=a(\nu){\rm d}t+b(\nu){\rm d}W.
\]
The set $(P,Q,\nu)$ evolves according to a slow-fast dynamics: the
angle variable $Q$ evolves on a timescale of order $\epsilon$, whereas the
variables $P$ and $\nu$ evolving on a timescale
of order one. Our aim is to average the system (\ref{eq:system slow-fast})
over the dynamics of $Q$ to obtain a closed system of equations describing
the dynamics of $\left(P,\nu\right)$. 

Before doing the averaging procedure, we first recall a very classical
result of Hamiltonian systems with slow time dependance (see e.g \cite{gardner1959adiabatic}
or \cite{lichtenberg2013regular} section 2.3). There exists a smooth
function $H_{1}\left(P,Q,\nu\right)$ such that the differential of
$(P,Q)$ with respect to $\nu$ can be expressed as 
\begin{equation}
\begin{cases}
\frac{\partial P}{\partial\nu} & =-\frac{\partial H_{1}}{\partial Q},\\
\frac{\partial Q}{\partial\nu} & =\frac{\partial H_{1}}{\partial P}.
\end{cases}\label{eq:canonical diff}
\end{equation}
We propose a simple proof of this result in Appendix \ref{sec:Canonical-change-of}
using differential two-forms. With the important result (\ref{eq:canonical diff}),
we can compute the second order differential of $P$ using the differentials
of $H_{1}$. We define the canonical Poisson brackets of any functions
$f(p,q)$ and $g(p,q)$ by
\[
\left\{ f,g\right\} _{p,q}:=\frac{\partial f}{\partial p}\frac{\partial g}{\partial q}-\frac{\partial f}{\partial q}\frac{\partial g}{\partial p}.
\]
We have 
\begin{align*}
\frac{\partial^{2}P}{\partial\nu^{2}} & =\frac{\partial}{\partial\nu}\left(-\frac{\partial H_{1}}{\partial Q}\right)\\
 & =-\frac{\partial^{2}H_{1}}{\partial P\partial Q}\frac{\partial P}{\partial\nu}-\frac{\partial^{2}H_{1}}{\partial Q^{2}}\frac{\partial Q}{\partial\nu}-\frac{\partial^{2}H_{1}}{\partial\nu\partial Q},
\end{align*}
and using once more (\ref{eq:canonical diff}) we obtain
\[
\frac{\partial^{2}P}{\partial\nu^{2}}=-\left\{ H_{1},\frac{\partial H_{1}}{\partial Q}\right\} _{P,Q}-\frac{\partial^{2}H_{1}}{\partial\nu\partial Q},
\]
with $\left\{ H_{1},\frac{\partial H_{1}}{\partial Q}\right\} _{P,Q}$
the canonical Poisson bracket with the set of variables $(P,Q)$.
A similar computation leads to 
\[
\frac{\partial^{2}Q}{\partial\nu^{2}}=\left\{ H_{1},\frac{\partial H_{1}}{\partial P}\right\} _{P,Q}+\frac{\partial^{2}H_{1}}{\partial\nu\partial P}.
\]

We write the slow-fast system of equations (\ref{eq:system slow-fast})
as 
\begin{equation}
\begin{cases}
{\rm d}Q & =\frac{1}{\epsilon}\omega(P,\nu){\rm d}t+\text{terms of order 0 in }\frac{1}{\epsilon},\\
{\rm d}P & =\left[-\frac{\partial H_{1}}{\partial Q}a(\nu)-\frac{1}{2}\left(\left\{ H_{1},\frac{\partial H_{1}}{\partial Q}\right\} _{P,Q}+\frac{\partial^{2}H_{1}}{\partial\nu\partial Q}\right)b^{2}(\nu)\right]{\rm d}t-\frac{\partial H_{1}}{\partial Q}b(\nu){\rm d}W,\\
{\rm d}\nu & =a(\nu){\rm d}t+b(\nu){\rm d}W.
\end{cases}\label{eq:good slow-fast}
\end{equation}

We are now is position of doing the averaging procedure for the system (\ref{eq:good slow-fast}). The averaging should be done with particular rules, as we are dealing with a stochastic system. There are two standard methods of averaging for such a stochastic system. The first one consists in writing the Fokker-Planck equation corresponding to (\ref{eq:good slow-fast}), and then find the limit of this equation when $\epsilon\rightarrow 0$ using an expansion in powers of $\epsilon$. This method is well described in the literature and can for example be found in \cite{mallick2002anomalous,bouchet2016large}. In the following, we use another fully equivalent method (in the spirit of the techniques used by \cite{freidlin1978averaging,freidlin1984random,kifer2004averaging}), in which we directly average the different terms in (\ref{eq:good slow-fast}).

We want to average  the dynamics of $P$ over the fast dynamics of the
angle $Q$. To leading order in $\frac{1}{\epsilon}$, the dynamics
of $Q$ is simply 
\[
\dot{Q}=\frac{1}{\epsilon}\omega(P,\nu).
\]
Therefore, the invariant measure of this dynamics, for any fixed values
of $P$ and $\nu$, is just the constant measure over the range $[0,2\pi]$.
To find the limit stochastic process for $P$ when $\epsilon$ goes
to zero, we have to average equation (\ref{eq:good slow-fast}) using
the invariant measure of $Q$. Some terms are very easy to compute.
Let $\left\langle .\right\rangle _{Q}:=\frac{1}{2\pi}\int_{0}^{2\pi}.~{\rm d}Q$
be the averaging operator over the fast dynamics of $Q$, we have
\begin{eqnarray*}
\left\langle \frac{\partial H_{1}}{\partial Q}\right\rangle _{Q}=0 & \text{and} & \left\langle \frac{\partial^{2}H_{1}}{\partial\nu\partial Q}\right\rangle _{Q}=0.
\end{eqnarray*}
In the deterministic part on the equation for $P$, the only nonzero
dependance comes from the average of the Poisson bracket $\left\langle \left\{ H_{1},\frac{\partial H_{1}}{\partial Q}\right\} \right\rangle _{Q}$. 

The average of the stochastic term $\frac{\partial H_{1}}{\partial Q}b(\nu){\rm d}W$ is more subtle, because it involves the stochastic process $W(t)$ that has variations on very small timescales. We detail the full procedure in Appendix \ref{sec:averaging}. The main idea is that the sum of independent Gaussian random variables is still a Gaussian random variable. As a consequence of this property, the average of the Gaussian white noise $\frac{\partial H_{1}}{\partial Q}b(\nu){\rm d}W$ is still a Gaussian white noise, because the average is composed of a sum of infinitely small Gaussian increments. For the standard Brownian motion for example, this property implies that $W(t)$ is self-similar, namely that for all $\tau>0$, $W(t/\tau)=(1/\sqrt{\tau})W(t)$. Because of the property of self-similarity, a diffusion process  is never slow in the strong sense. It can be considered as "slow" in the sense that its probability density evolves on a slow timescale. More details are given in Appendix \ref{sec:averaging}.

To average over
$Q$, we introduce first the matrix
\[
\sigma:=\begin{pmatrix}b(\nu)\\
b(\nu)\frac{\partial H_{1}}{\partial Q}
\end{pmatrix}.
\]
The noise term acting on the set of slow variables $(\nu,P)$ can
be written as $\sigma{\rm d}W$, where $W$ is a Wiener process. We
then use the standard result that the noise term
\[
\sigma\left(Q\left(\frac{t}{\epsilon}\right)\right){\rm d}W
\]
is equivalent (for the probability distribution) when $\epsilon\rightarrow0$
to a Gaussian white noise process, the variance of which is given
by the averaged correlation matrix $\left\langle \sigma\sigma^{T}\right\rangle _{Q}$(see
e.g \cite{freidlin1984random} chapter 8, and Appendix \ref{sec:averaging}). The correlation matrix
of the noise is
\[
\sigma\sigma^{T}=\begin{pmatrix}b^{2}(\nu) & b^{2}(\nu)\frac{\partial H_{1}}{\partial Q}\\
b^{2}(\nu)\frac{\partial H_{1}}{\partial Q} & b^{2}(\nu)\left(\frac{\partial H_{1}}{\partial Q}\right)^{2}
\end{pmatrix},
\]
and averaging over $Q$ gives
\begin{equation}
\left\langle \sigma\sigma^{T}\right\rangle _{Q}=\begin{pmatrix}b^{2}(\nu) & 0\\
0 & b^{2}(\nu)\left\langle \left(\frac{\partial H_{1}}{\partial Q}\right)^{2}\right\rangle _{Q}
\end{pmatrix}.\label{eq:average correlations}
\end{equation}
The important consequence of (\ref{eq:average correlations}) is that
the average over the fast dynamics eliminates the correlations between
the noise terms in the equations of $P$ and $\nu$. 

We can now present our main theoretical result. The dynamics of the
slow process $(P,\nu)$ follows when $\epsilon$ goes to zero the
averaged equations
\begin{equation}
\begin{cases}
{\rm d}P & =-\frac{1}{2}b^{2}(\nu)\left\langle \left\{ H_{1},\frac{\partial H_{1}}{\partial Q}\right\} \right\rangle _{Q}{\rm d}t+b(\nu)\sqrt{\left\langle \left(\frac{\partial H_{1}}{\partial Q}\right)^{2}\right\rangle _{Q}}{\rm d}W_{1},\\
{\rm d}\nu & =a(\nu){\rm d}t+b(\nu){\rm d}W_{2}.
\end{cases}\label{eq:average action}
\end{equation}
There are some interesting comments to do on equations (\ref{eq:average action}).
\begin{enumerate}
\item First, we note that the Wiener processes $W_{1}$ and $W_{2}$ involved
in the two equations for $P$ and $\nu$ are independent. This beautiful
property is a consequence of the relation $\frac{\partial P}{\partial\nu}=-\frac{\partial H_{1}}{\partial Q}.$
It seems counterintuitive that two independent Wiener processes can appear as the limit of a single Wiener process. We can heuristically explain this result as follows. A large "kick" created by the Wiener process is in fact the cumulative result of a large number of small "kicks" (see \ref{sec:averaging} for a more detailed description). But because of the presence of the amplitude $\frac{\partial H_{1}}{\partial Q}$ in the equation for $P$, each "kick" is weighted by a particular value of $\frac{\partial H_{1}}{\partial Q}$,  and the sum is composed of terms that cancel each other. A large kick in the equation for $\nu$ is not felt by the action $P$ because the cumulative effect is destroyed by the weight function $\frac{\partial H_{1}}{\partial Q}$. Therefore, noise-stimulated events for $\nu$ and $P$ are related to completely different realizations of the Wiener process.
It would not be  the case if we had chosen to study the dynamics of
another slow variable of the system, for example the Hamiltonian $H$.
\item Also, we observe that the diffusion of the action $P$ only comes
from the stochastic part of the equation for $\nu$, the coefficient
$b(\nu)$. For a smooth function $\nu(t)$, the action cannot diffuse.
This is in accordance with the theory of adiabatic invariants, that
states that for a smooth time dependance of $\nu(\epsilon t)$, there
exists an adiabatic invariant conserved to any order in $\epsilon$
(This result was found by \cite{kulsrud1957adiabatic} for the harmonic
oscillator, and \cite{gardner1959adiabatic} in the general case).
In the case where such an invariant exists, the action cannot diffuse.
\end {enumerate}

An other interesting remark is that the stochastic equation (\ref{eq:average action})
for $P$ can be written in a more compact way. From the relation
\[
\frac{\partial}{\partial Q}\left(\frac{\partial H_{1}}{\partial P}\frac{\partial H_{1}}{\partial Q}\right)=\frac{\partial H_{1}}{\partial P}\frac{\partial^{2}H_{1}}{\partial Q^{2}}+\frac{\partial^{2}H_{1}}{\partial Q\partial P}\frac{\partial H_{1}}{\partial Q}
\]
averaged over $Q$, we get
\begin{equation}
\left\langle \frac{\partial H_{1}}{\partial P}\frac{\partial^{2}H_{1}}{\partial Q^{2}}\right\rangle _{Q}=-\left\langle \frac{\partial^{2}H_{1}}{\partial Q\partial P}\frac{\partial H_{1}}{\partial Q}\right\rangle _{Q}.\label{eq:relationX}
\end{equation}
Using (\ref{eq:relationX}), equation (\ref{eq:average action}) for
$P$ can be equivalently written
\[
{\rm d}P=b^{2}(\nu)\frac{\partial}{\partial P}\left\langle \left(\frac{\partial H_{1}}{\partial Q}\right)^{2}\right\rangle _{Q}{\rm d}t+b(\nu)\sqrt{\left\langle \left(\frac{\partial H_{1}}{\partial Q}\right)^{2}\right\rangle _{Q}}{\rm d}W_{1}.
\]
Interestingly, the last equation shows that the drift and the noise amplitude are related to the same function, and that one can be obtained from the other. 

In the following section \ref{subsec:application-oscillator},
we present a simple applications of the theoretical result (\ref{eq:average action}).
We study the well-known harmonic oscillator with random frequency, which allows for explicit computations.

\subsection{Application to the harmonic oscillator\label{subsec:application-oscillator}}

The Hamiltonian of the harmonic oscillator is 
\begin{equation}
H(p,q,\nu)=\frac{1}{2}p^{2}+\frac{1}{2}\nu^{2}q^{2}.\label{eq:hamiltonian oscillator}
\end{equation}
and the set
of Hamilton's equations (\ref{eq:hamiltonien modele}) that describes the dynamics of the canonical
variables $(p,q)$ on the slow timescale
is 
\begin{equation}
\begin{cases}
\dot{p} & =-\frac{1}{\epsilon}\frac{\partial H}{\partial q}\left(p,q,\nu(t)\right)\\
\dot{q} & =\frac{1}{\epsilon}\frac{\partial H}{\partial p}\left(p,q,\nu(t)\right)
\end{cases}\label{eq:Hamilton_oscillator}
\end{equation}
We consider that the frequency of the oscillator $\nu$ is a parameter
in the Hamiltonian (\ref{eq:hamiltonian oscillator}), which time
evolution is given by the It\^o stochastic differential equation
\begin{equation}
{\rm d}\nu=a(\nu){\rm d}t+b(\nu){\rm d}W.\label{eq:random frequency}
\end{equation}
The small parameter $\epsilon\ll1$ sets the timescale separation
between the Hamiltonian dynamics, and the dynamics of the random frequency.
Equations (\ref{eq:Hamilton_oscillator}-\ref{eq:random frequency})
define our model.

We introduce the classical action-angle variables $(P,Q)$ defined
by the relations
\begin{align}
p & =\sqrt{2\nu P}\cos Q,\nonumber \\
q & =\sqrt{\frac{2P}{\nu}}\sin Q,\label{eq:canonical variables}
\end{align}
and the new Hamiltonian $\widetilde{H}(P,\omega)$ simply becomes
\[
\widetilde{H}(P,\nu)=\nu P.
\]
To use the result (\ref{eq:average action}), we have to find the
expression of $H_{1}$ as a function of $P,Q$ and $\nu$. From the
expression $P=\frac{p^{2}}{2\nu}+\frac{\nu q^{2}}{2}$ we compute
\begin{align*}
\frac{\partial P}{\partial\nu} & =-\frac{1}{2\nu^{2}}p^{2}+\frac{1}{2}q^{2}\\
 & =-\frac{1}{\nu}P\cos^{2}Q+\frac{1}{\nu}P\sin^{2}Q\\
 & =-\frac{\partial}{\partial Q}\left[\frac{1}{2\nu}P\sin\left(2Q\right)\right].
\end{align*}
A straightforward calculation also shows that 
\begin{align*}
\frac{\partial Q}{\partial\nu} & =\frac{1}{2\nu}\sin\left(2Q\right)\\
 & =\frac{\partial}{\partial P}\left[\frac{1}{2\nu}P\sin\left(2Q\right)\right].
\end{align*}
This gives the function $H_{1}(P,Q,\nu)$ (\cite{lichtenberg2013regular}
section 2.3)
\[
H_{1}=\frac{1}{2\nu}P\sin\left(2Q\right).
\]
In the case of the harmonic oscillator, equations (\ref{eq:average action})
are explicitly 
\begin{equation}
\begin{cases}
{\rm d}P & =b^{2}(\nu)\frac{P}{2\nu^{2}}{\rm d}t+b(\nu)\frac{P}{\sqrt{2}\nu}{\rm d}W_{1},\\
{\rm d}\nu & =a(\nu){\rm d}t+b(\nu){\rm d}W_{2}.
\end{cases}\label{eq:slow action oscillator}
\end{equation}

In order to illustrate the result (\ref{eq:slow action oscillator}),
we perform a numerical simulation. We have to choose a stochastic
process. We choose a process such that the frequency
$\nu$ is always strictly positive, otherwise action variables are ill-defined. We thus set $b(\nu)=\sqrt{2\sigma^{2}}$,
where $\sigma$ is a constant parameter, and $a(\nu)=-\nabla V(\nu)$,
where the potential $V(\nu)=\frac{1}{\nu}+\frac{1}{\nu_{max}-\nu}$
is chosen such that the frequency $\nu$ is trapped in the range $]0,\nu_{max}[$.

\begin{figure}
\includegraphics[height=5cm]{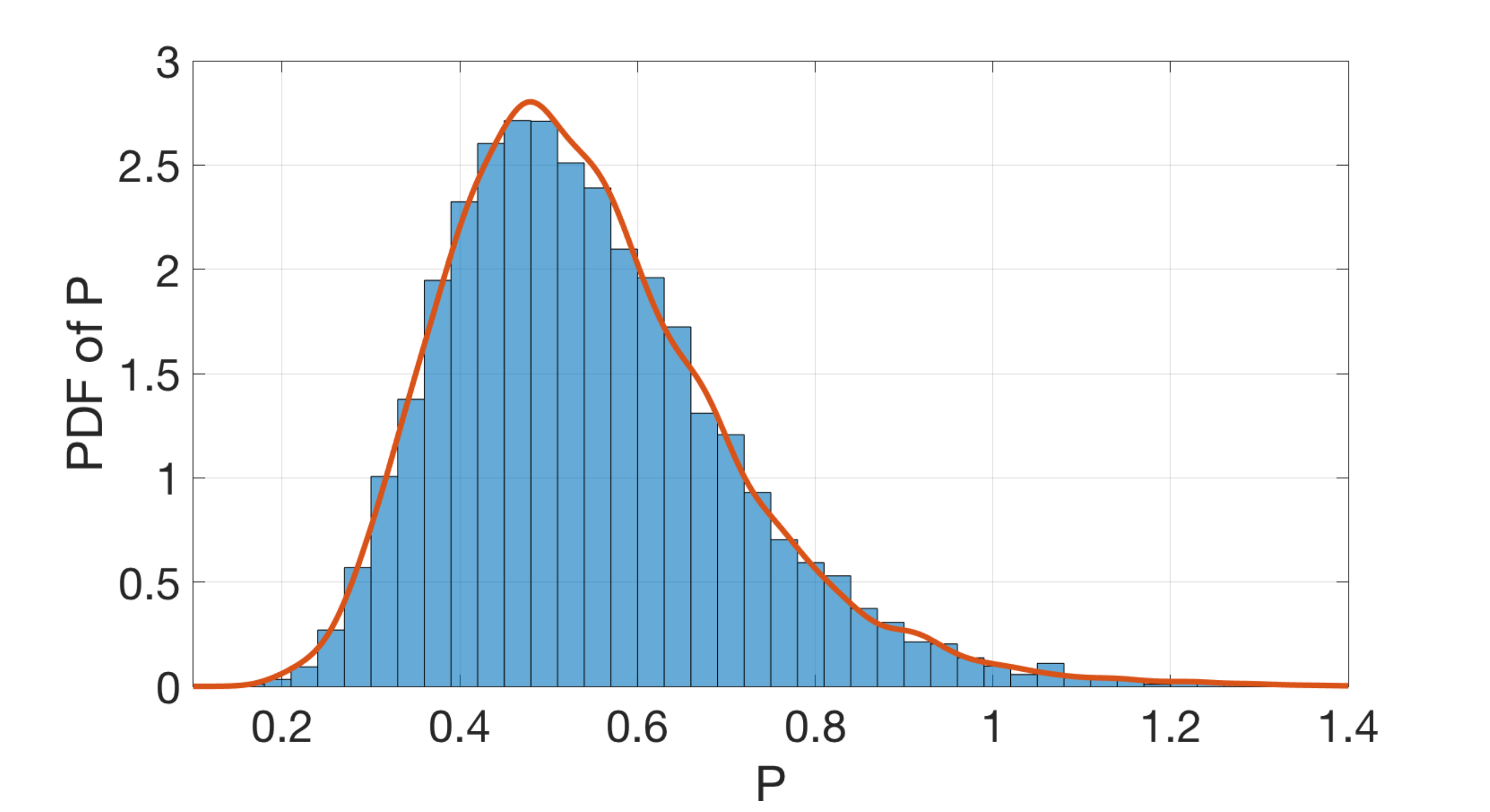}\includegraphics[height=5cm]{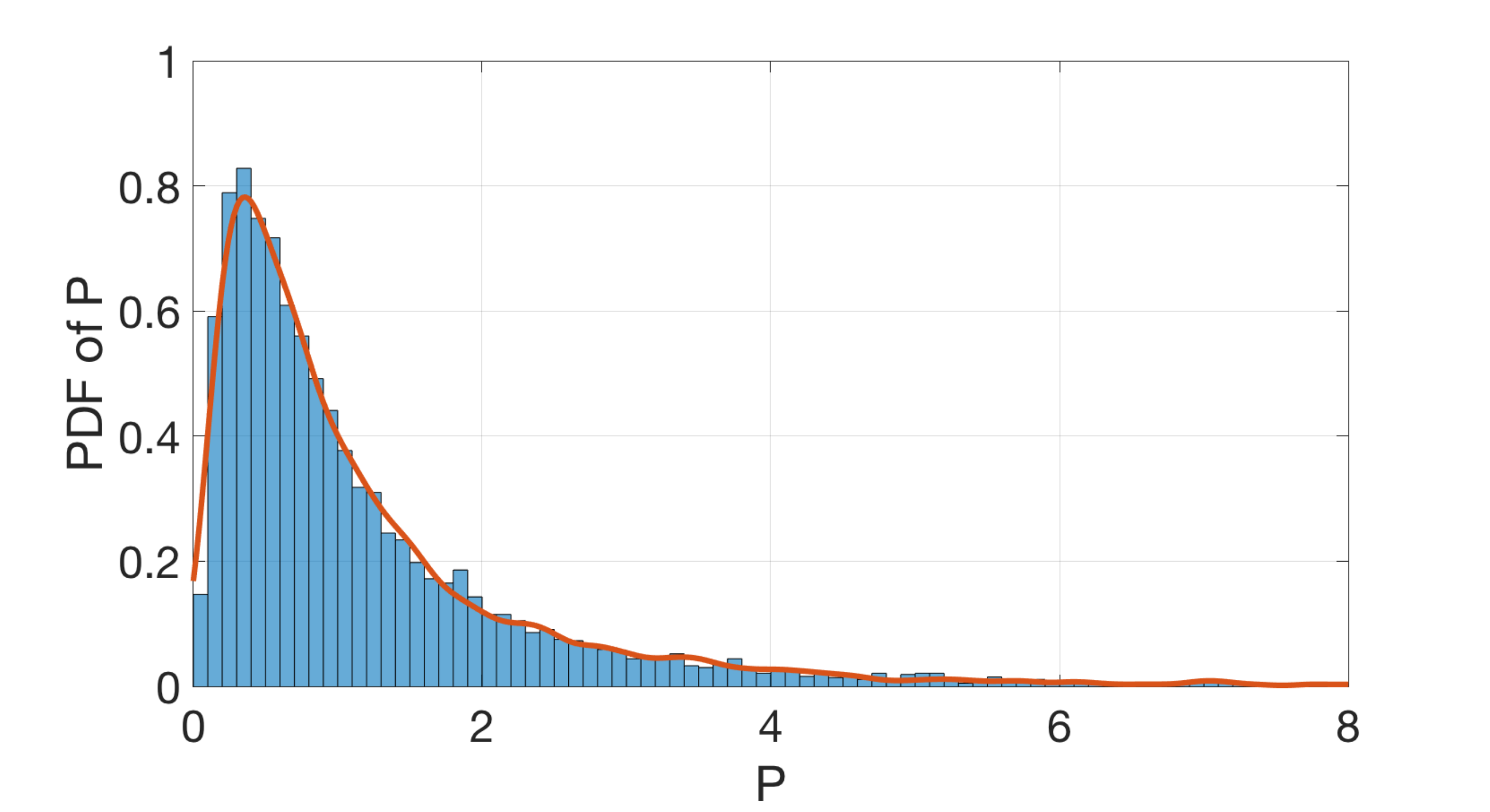}

\caption{PDF of the canonical action $P$ obtained either with Hamilton's equations
(histogram) or with the averaged equations (curve), at $t=50$, and
$t=500$. \label{fig:result oscillator}}

\end{figure}

We integrate Hamilton's equations of motion
\[
\begin{cases}
\dot{q} & =\frac{1}{\epsilon}p,\\
\dot{p} & =-\frac{\nu^{2}}{\epsilon}q,
\end{cases}
\]
using a symplectic integrator of order 2. And we integrate simultaneously
the stochastic equation for $\nu$
\[
{\rm d}\nu=\left[\frac{1}{\nu^{2}}-\frac{1}{\left(\nu_{max}-\nu\right)^{2}}\right]{\rm d}t+\sqrt{2\sigma^{2}}{\rm d}W,
\]
using a stochastic Euler algorithm (described e.g. in \cite{gardiner1985stochastic} chapter 15). The integration is done over $10,000$
realizations of the stochastic frequency, and the same initial conditions
$(p_{0},q_{0})=(1,0)$, $\nu_{0}=\frac{\nu_{max}}{2}$ for each trajectory.
The parameters are $\nu_{max}=2.0$, $\sigma=0.3$ and $\epsilon=0.01$.
The histogram of the action $P=\frac{p^{2}}{2\nu}+\frac{\nu q^{2}}{2}$
is represented at different times on figure (\ref{fig:result oscillator})
by the histograms.

Secondly, we perform $10,000$ integrations of the averaged equations
\[
\begin{cases}
{\rm d}P & =\sigma^{2}\frac{P}{\nu^{2}}{\rm d}t+\sigma\frac{P}{\nu}{\rm d}W_{1},\\
{\rm d}\nu & =\left[\frac{1}{\nu^{2}}-\frac{1}{\left(\nu_{max}-\nu\right)^{2}}\right]{\rm d}t+\sqrt{2\sigma^{2}}{\rm d}W_{2},
\end{cases}
\]
with the same parameters and initial conditions. The histogram of
the action is represented by the curve on figure (\ref{fig:result oscillator}).
Both agree up to sampling errors, as expected. The main interest of
using the averaged equations to compute a PDF is of course the drastic
reduction of the time of integration. Moreover, one can then recover
the PDF of the initial variables $(p,q)$ using the change of variable
and the fact that the PDF of $Q$ is uniform over $[0,2\pi]$. 

\section{The chaotic case\label{sec:The-chaotic-case}}

\subsection{Formulation of the model\label{subsec:Formulation-of-the}}

We now turn to a more difficult situation. We study the dynamics (\ref{eq:hamiltonien modele})
where the Hamiltonian $H(x,\nu)$ has one and a half degrees of freedom. More precisely, the set of canonical variables is $\{(p,q),(\Lambda,\lambda)\}$ but $\Lambda$ does not play any role in the dynamics.
We assume that the explicit time dependance in $H$ creates many resonances
overlap, and thus the Hamiltonian dynamics for fixed $\nu$ has both
regular and chaotic trajectories. The phase space is divided into
regions with different mixing properties. In the regions in $(p,q)$-space
where at least two resonances overlap, the dynamics is strongly chaotic.
In regions of phase space that are far away from the main resonances,
the dynamics is almost regular. KAM-tori may subsist or not, depending
on the amplitude and localization of the resonant frequencies. The
parameter $\nu$ is assumed to be stochastic, and evolving on a time
scale of order one, whereas the canonical variables evolve on the
small timescale $\epsilon\ll1$. We want to characterize transport
in phase space on the slow timescale.

By contrast with the simplest integrable case of section \ref{sec:The-integrable-case},
there is now no global action variable. The transport cannot be described
by a diffusive equation of some kind of action variable. What we will
explain in the present section is that the transport is very different
in the present case where the dynamics of $H$ is chaotic. In the
chaotic case, there are two competitive transport mechanisms, that
can be qualitatively described as follows:
\begin{enumerate}
\item The chaotic regions are moving in phase space. If the system is in
one of those chaotic regions, it is carried together with the region
upwards or downwards. At any time, it can leave the region and enter
in the regular part of phase space. But depending on when the system
leaves the region, it may be carried up or down far away from its
initial position. We refer to this mechanism as ``transport by migration
with chaotic regions''.
\item The regular regions far away from the resonances are very similar
to the integrable Hamiltonian dynamics of section \ref{sec:The-integrable-case}.
The slow stochastic variations of $\nu$ distort the orbits, and enhances
the chaotic diffusion of these regions. As a result, no KAM tori can
subsist and the system diffuses through phase space on a time scale
of order one. We refer to this mechanism as ``noise driven transport
in regular regions''.
\end{enumerate}
Depending on the parameters of the Hamiltonian, one of the two mechanisms
described above may overcome the other. This section provides a concrete
example of a Hamiltonian where the two mechanisms of transport are
present. We also determine in which parameter regime the transport
by migration with chaotic regions is dominant. 

We propose to study the Hamiltonian dynamics (\ref{eq:hamiltonien modele}),
where the Hamiltonian depends on a set of four frequencies
\begin{equation}
H(p,q,\Lambda,\lambda,\nu)=\frac{p^{2}}{2}+\stackrel[k=1]{4}{\sum}\cos\left(q-\lambda_k-\varphi_{k}\right)+\stackrel[k=1]{4}{\sum}\nu_k\Lambda_k,\label{eq:toy hamiltonian}
\end{equation}
Where $\{\Lambda,\lambda\}$ are now conjugated canonical variables of dimension four. The dynamics (\ref{eq:hamiltonien modele}) is explicitly
\begin{equation}
\begin{cases}
\dot{q} & =\frac{1}{\epsilon}p\\
\dot{p} & =\frac{1}{\epsilon}\stackrel[k=1]{4}{\sum}\sin\left(q-\lambda_k-\varphi_{k}\right)\\
\dot{\lambda} & =\frac{1}{\epsilon}\nu
\end{cases}\label{eq:canonic chaotic}
\end{equation}
The canonical action $\Lambda$ does not play any role in the dynamics. The set of four frequencies $\nu:=(\nu_{1},\nu_{2},\nu_{3},\nu_{4})$
plays the role of the external parameters. The frequencies are divided
in two groups of two frequencies which create resonance overlap, and
create two main chaotic regions around $p=0$ and $p=10$ respectively.
The parameters $\left\{ \varphi_{k}\right\} _{k=1..4}$ are some initial
phases, and $\epsilon$ is a small parameter to model the timescale
separation between the fast dynamics of the canonical variables and
the stochastic dynamics of the frequencies. To complete our Hamiltonian
model (\ref{eq:toy hamiltonian}), we need to specify the stochastic
process for the set of frequencies $\nu$. We choose for the variations
of $\nu$ an Ornstein-Uhlenbeck process defined by
\begin{equation}
{\rm d}\nu=-(\nu-\nu^{*}){\rm d}t+\sqrt{2\sigma^{2}}{\rm d}W.\label{eq:OU process}
\end{equation}
In Equation (\ref{eq:OU process}), $W(t)$ is a 4-dimensional Wiener
process. In order to keep the size of the stochastic
regions constant, we choose to prescribe the same noise for two frequencies
of the same set, that is, $W_{1}=W_{2}$ and $W_{3}=W_{4}$. The correlation
function of the noise is $\left\langle {\rm d}W_{1}(t){\rm d}W_{3}(t')\right\rangle =0$
and $\left\langle {\rm d}W_{1}(t){\rm d}W_{1}(t')\right\rangle =\left\langle {\rm d}W_{3}(t){\rm d}W_{3}(t')\right\rangle =\delta(t-t'){\rm d}t$.
The parameter $\sigma$ quantifies the noise amplitude and is the
same for all frequencies. The term $-(\nu-\nu^{*})$ keeps the frequencies
close to their averaged values defined by the set $\nu^{*}$. 
The choice to keep pairs of frequencies with the same noise has no physical justification, it is only a simplifying assumption to keep constant the widths of the chaotic regions. In the general case, we expect both the width and the position of the chaotic regions to vary with time.  If the frequencies of a pair do not have the same noise, a mechanism of transport by slow extension of the chaotic regions occurs. However, this third mechanism is very similar to the mechanism of transport through slow migration of chaotic regions, and can be accounted for using the same techniques as the ones presented in this section. An extension of the theory to the general case is left apart for further work.\\

The Hamiltonian (\ref{eq:toy hamiltonian}) together with the stochastic
equation (\ref{eq:OU process}) completely defines our model. In the
simulations of the dynamics (\ref{eq:canonic chaotic}), the parameters $\varphi_{k}$, the timescale
separation $\epsilon$ and the mean frequencies $\nu_{k}^{*}$ were
fixed to the values given in table (\ref{tab:parameters}), whereas
the amplitude $\sigma$ is a control parameter that we changed in
the different simulations.

\begin{figure}
\begin{centering}
\includegraphics[height=5cm]{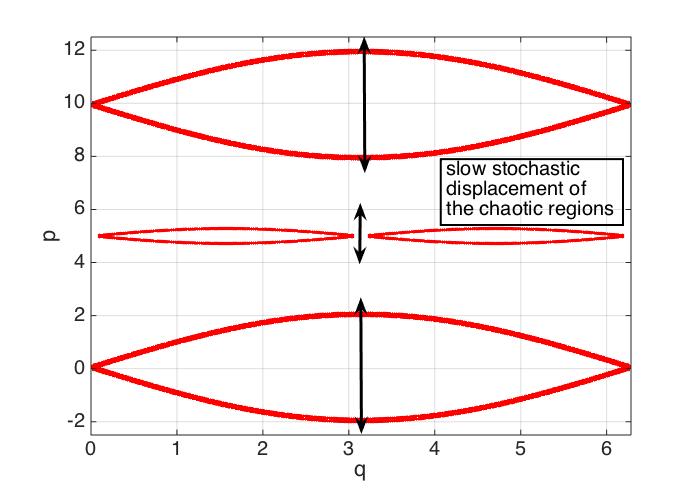}\includegraphics[height=5cm]{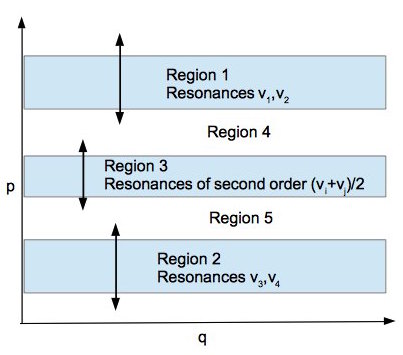}
\par\end{centering}
\caption{Schematic representation of the chaotic structure of phase space for
the Hamiltonian (\ref{eq:toy hamiltonian}). The left panel represents
The eyes of the resonances of first and second order in the system.
Resonance overlap creates strongly chaotic regions in phase space.
The chaotic regions are represented by the bands on the right panel.
The arrows indicate that the chaotic regions are slowly moving with
time. \label{fig:schema_model}}
\end{figure}

\begin{table}
\begin{centering}
\begin{tabular}{|c|c|c|}
\hline 
frequencies $\nu_{k}^{*}$ ($s^{-1}$) & initial angles $\varphi_{k}$  & timescale separation $\epsilon$\tabularnewline
\hline 
\hline 
10.0 & 0.0 & \tabularnewline
\cline{1-2} 
9.9 & $\pi$ & $10^{-2}$\tabularnewline
\cline{1-2} 
0.1 & $\pi$ & \tabularnewline
\cline{1-2} 
0.0 & 0.0 & \tabularnewline
\hline 
\end{tabular}
\par\end{centering}
\caption{Fixed parameters of the model (\ref{eq:toy hamiltonian}-\ref{eq:OU process})\label{tab:parameters}}
\end{table}
In figure (\ref{fig:schema_model}), we have represented two pictures
to help the reader understand the chaotic structure of phase space.
On the left, we have represented the cat's eyes of the main resonances
in red. All the regions of phase space covered by the eyes are strongly
chaotic. The two first order resonances appear with amplitude 1 in
the Hamiltonian (\ref{eq:toy hamiltonian}), and thus create cat's
eyes  with extension 4. The second order resonances create a chaotic
region around $p=5$, and its extension can be found using Lie transformations
of the Hamiltonian (\ref{eq:toy hamiltonian}). We found that the
second order resonances come with typical amplitude $\frac{2}{\nu_{1}^{2}}$
and have thus an extension $4\sqrt{\frac{2}{\nu_{1}^{2}}}\approx0.56.$
Those three chaotic regions are separated by two regular bands, in
which transport is only possible either through the weak chaos created
by higher order resonances, that are of much smaller amplitudes, or
through noise driven transport in regular regions. On the right of
figure (\ref{fig:schema_model}), we have represented the bands associated
with the chaotic regions of first and second orders. For convenience,
we have labelled the regions. Regions 1 and 2 are the chaotic regions
of first order resonances, and region 3 is the chaotic region of second
order resonances. The regions of weak chaos, which we also call the
``regular'' regions, are labelled as regions 4 and 5. Within the
bands 1, 2 and 3, chaos is strongly developed, and the system is thus
rapidly carried throughout the band on the timescale $\epsilon$.
The bands are separated by regions 4 and 5 of much weaker chaos: for
fixed values of the frequencies $\nu$, the system can hardly cross
those regular regions, and therefore the migration between one band
to the other is very slow. \\

The important point to emphasize is that figure (\ref{fig:schema_model})
is not static: the set of four frequencies $\nu$ is slowly varying
around its mean values $\nu^{*}$, with stochastic variations defined
by equation (\ref{eq:OU process}). This slow variation of the frequencies
imply that the transport in phase space does not only happen through
the well known \textquotedbl chaotic diffusion\textquotedbl{} observed
in chaotic maps (e.g the standard map, see \cite{lichtenberg2013regular}
chapter 5). It comes from two other mechanisms that we have previously
referred to as \textquotedbl transport by migration with chaotic
regions\textquotedbl{} and ``noise driven transport in regular regions''.
We look in section \ref{subsec:Numerical-simulations} at the problem
of first passage time at $p=0$ starting at $p=10$. We will see that
the first mechanism, that is, transport by migration with the chaotic
regions, is dominant for large noise amplitude.

\begin{figure}
\centering{}\includegraphics[width=16cm]{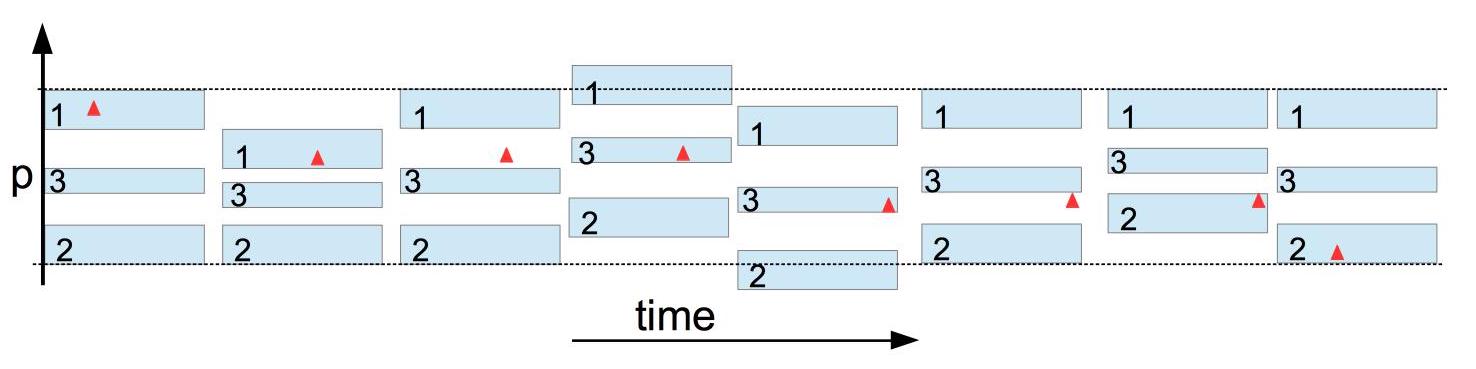}\caption{Schematic representation of a trajectory in the Hamiltonian stochastic
model (\ref{eq:toy hamiltonian}). The system is represented by the
red triangle.\label{fig:trajectory example} The picture displays
the chaotic regions 1, 2 and 3 at 8 different times. The system is
initially located in region 1 and is carried to region 2 through the
mechanism of transport by migration of the chaotic regions.}
\end{figure}

\begin{figure}
\includegraphics[height=6cm]{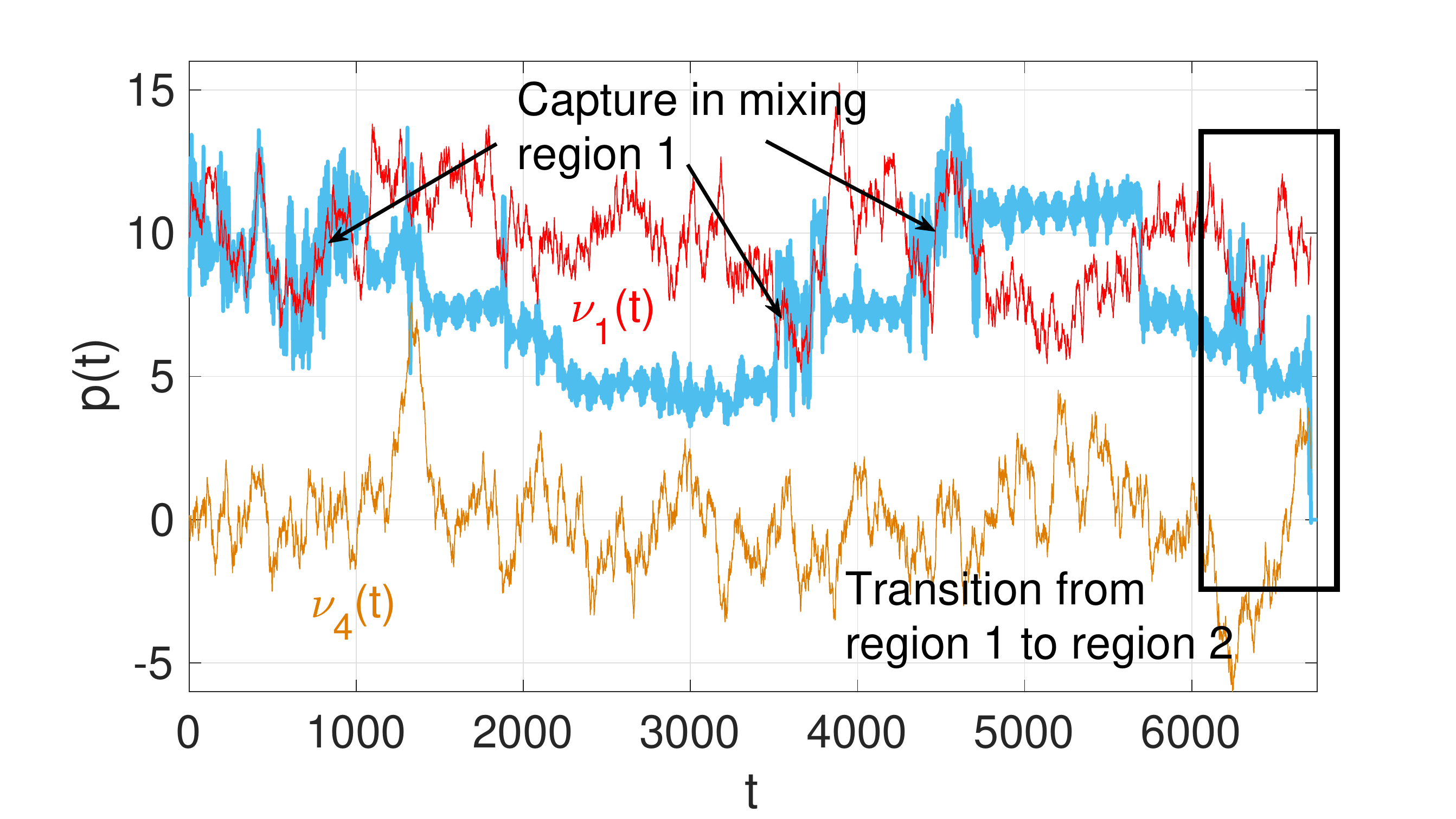}\includegraphics[height=6cm]{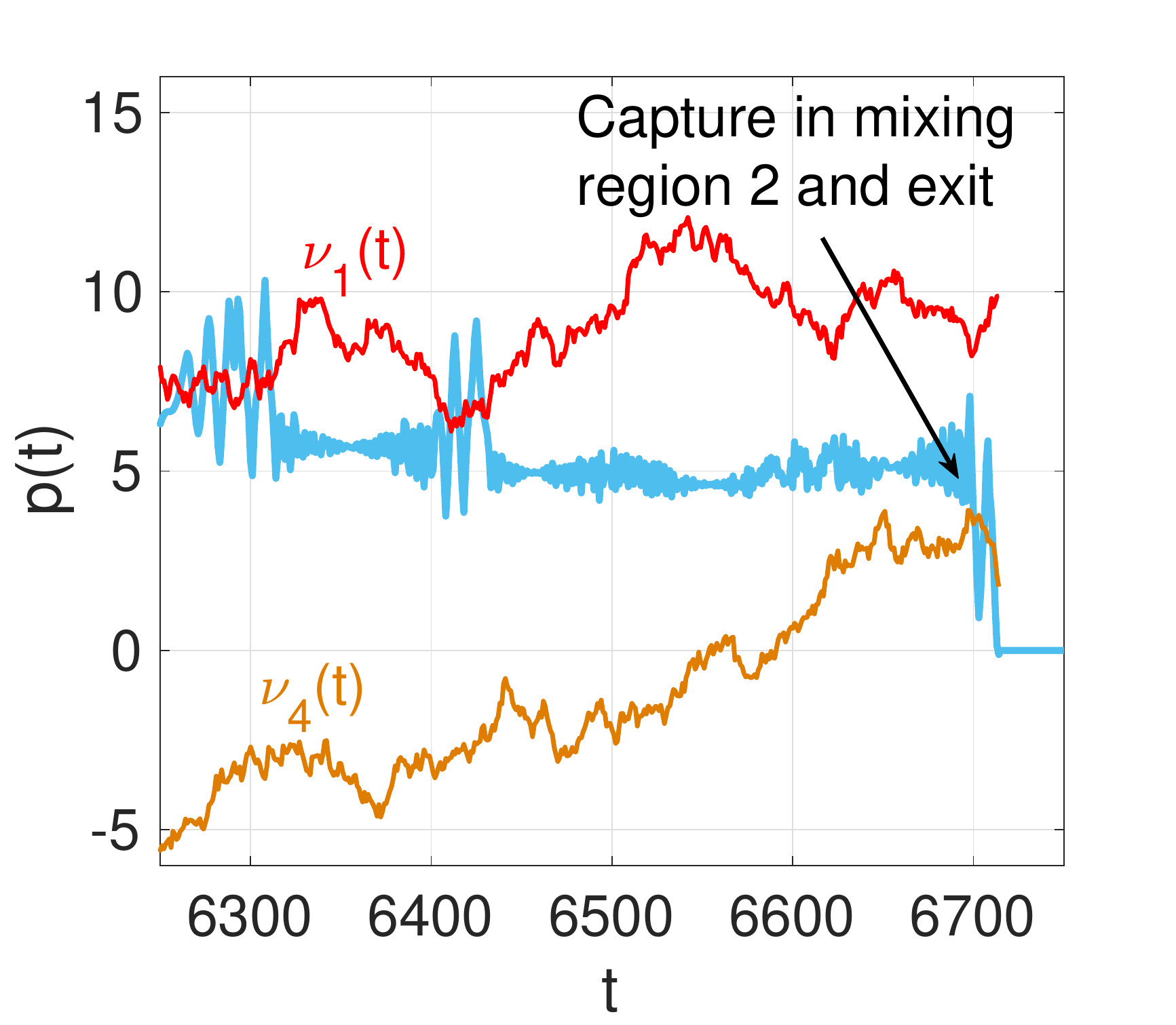}

\caption{Example of a trajectory in the stochastic Hamiltonian model (\ref{eq:toy hamiltonian})
transported from $p=10$ to $p=0$ by the stochastic migration of
the chaotic regions. The blue curve is the action $p(t)$. The red
and orange curves represent the frequencies $\nu_{1}(t)$ and $\nu_{4}(t)$.
When the system is located in the chaotic region 1, the action value
has fast and large fluctuations around $\nu_{1}(t)$. When the system
leaves region 1 and enters in the weakly chaotic region 4, the action
fluctuations are much smaller. The left panel is an enlargement of
the end of the trajectory, when the system is captured by the chaotic
region 2 and exits through the boundary at $p=0$. \label{fig:ex out_trajectory}}

\end{figure}

Let us assume for example that we are in a range of parameters for
which the transport is mainly due to migration with the chaotic regions.
The system starts at $p=10$ . It can reach the value $p=0$ through
successive jumps from one region to the other, until it eventually
reaches the chaotic region around the frequency $\nu_{4}^{*}=0$.
To illustrate the transport mechanism, we have represented in figure
(\ref{fig:trajectory example}) the different steps of the transport.
We have labelled the chaotic and regular regions the same way as in
figure (\ref{fig:schema_model}). First, one downward fluctuation
of the couple of frequencies $(\nu_{1},\nu_{2})$ brings together
the chaotic region 1 and the system around $p=7$. Then, region 1
moves upward again, but the system leaves the chaotic region and is
thus trapped in region 4. A simultaneous displacement upwards of the
frequencies $\nu_{3}$ and $\nu_{4}$ brings the region 3 of second
order resonances upwards, and it captures the system. The system has
thus passed from region 1 to region 3 thanks to stochastic migration
with the chaotic regions. It then passes from region 3 to region 2
by a similar trajectory mechanism: it is transported downwards and
left in region 5 and an eventual upward displacement of region 2 captures
it and brings it to $p=0$. Figure (\ref{fig:ex out_trajectory})
displays an example of a trajectory starting at $p=10$ and reaching
$p=0$. The example of figure (\ref{fig:ex out_trajectory}) shows
that transport is not straightforward to $p=0$. The system can be
captured and released many times by a chaotic region before being
captured by another chaotic region. In the example of figure (\ref{fig:ex out_trajectory}),
the system spends most of the time in the vicinity of region 1, and
the transition to regions 3 and 2 only occurs at the very end of the
trajectory.\\

The mechanism of transport we just described, composed of successive
jumps between regions of different types, is typical to go from $p=10$
to $p=0$ when chaotic diffusion in the regular regions is negligible.
As the reader would surely have noticed, it is not important where
the system is exactly located when it is inside a chaotic region of
type 1, 2 or 3. The mixing in those strongly chaotic regions is so
fast compared to the timescale of frequency variations, that the system
has time to explore the whole region before any significant displacement
of the region. To say it another way, the underlying Hamiltonian dynamics
is not important to determine the transport characteristics. In fact,
there are only two properties of the dynamics that matter. The first
one is the conservation of area which is characteristic of Hamiltonian
dynamics (the Hamiltonian flow is a symplectic transformation). The
second one is that the phase space is partitioned into several strongly
mixing regions (where chaos is strongly developed), separated by regular
regions. Those two properties prompted us to perform a kind of ``averaging''
of the Hamiltonian dynamics and build an even simpler, fully stochastic
model, that we call the \emph{local diffusion model}. 


\subsection{Averaging of the dynamics: the local diffusion model\label{subsec:averaging-of-the}}

The local diffusion model is a purely stochastic model built from
the Hamiltonian model (\ref{eq:toy hamiltonian}) with stochastic
frequencies (\ref{eq:OU process}). The idea is to average the dynamics
over an intermediate timescale which is much longer than the timescale
of the Hamiltonian dynamics, but much smaller than the timescale of
stochastic variations of the frequencies. We use the hypothesis that
the regions of strong chaos are also mixing. As the timescale of the
Hamiltonian dynamics is typically of order $\frac{1}{\nu_{1}^{*}}$,
the averaging procedure is done over a time $\tau_{av}$ satisfying
\begin{equation}
\frac{\epsilon}{\nu_{1}^{*}}\ll\tau_{av}\ll\frac{1}{\nu_{1}^{*}}.\label{eq:timescale separation}
\end{equation}
Over the timescale $\tau_{av}$, the Hamiltonian dynamics in the chaotic
regions is assumed to be mixing. Strong chaos separates the neighboring
trajectories exponentially fast, on a timescale of order $\frac{\epsilon}{\nu_{1}^{*}}$,
and the system has thus completely ``forgotten'' its initial condition
over the timescale $\tau_{av}$. This means that if the system has
an initial condition inside the chaotic region 1 of figure (\ref{fig:trajectory example})
(for example), it can be anywhere inside the region 1 after a time
$\tau_{av}$. On the other hand, we assume that chaos is weak enough
in the regular regions 4 and 5 of figure (\ref{fig:trajectory example})
to keep the system to an average value very close to $p$ within the
time $\tau_{av}$.

\begin{figure}
\begin{centering}
\includegraphics[scale=0.5]{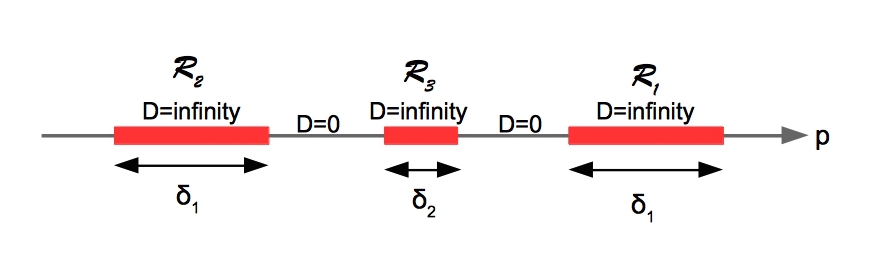}
\par\end{centering}
\caption{Schematic representation of the local diffusion model. The regions
1, 2 and 3 with infinite diffusion coefficients are displayed by the
red rectangles.\label{fig:local diffusive}}
\end{figure}
The local diffusion model of second order is represented in figure
(\ref{fig:local diffusive}). It consists of three patches of infinite
diffusion coefficient $D$ in $p$-space. Two of them have an extension
$\delta_{1}$, and correspond in figure (\ref{fig:schema_model})
to the chaotic regions 1 and 2 of main resonances. The third one has
a smaller extension $\delta_{2}$ and corresponds to region 3 of second
order resonances in figure (\ref{fig:schema_model}). We assume that
chaotic transport is negligible away from the resonances of first
and second orders, that's why we set the diffusion coefficient to
zero out of the diffusion patches. The diffusion patches are then
moved according to the stochastic dynamics (\ref{eq:OU process}). 

The solution cannot be represented any more by a trajectory, but only
through the probability distribution $\rho_{\nu}(p,t)$ to find the
system at impulsion $p$ at time $t$, given a realization of the
stochastic process $\nu(t)$. Let $\delta_{i}$ be the extension of
the i-th diffusion region that we call $\mathcal{R}_{i}$. The diffusion
region $\mathcal{R}_{i}$ then covers the interval $[\nu_{i}-\delta_{i}/2,\nu_{i}+\delta_{i}/2]$.
If the impulsion $p$ is out of all the diffusion regions, the function
$\rho_{\nu}(p,t)$ remains the same at step $t+{\rm d}t$. If $p$
is inside the diffusion region $\mathcal{R}_{i}$, the probability
distribution at step $t+{\rm d}t$ is the average of the probability
distribution over the whole region. The dynamics of the distribution
$\rho_{\nu}(p,t)$ can be implemented following the equations
\begin{align}
\rho_{\nu}(p,t+{\rm d}t) & =\begin{cases}
\frac{1}{\delta_{i}}\int_{\nu_{i}(t)-\delta_{i}/2}^{\nu_{i}(t)+\delta_{i}/2}\rho_{\nu}(p',t){\rm d}p' & \text{if p is in region \ensuremath{\mathcal{R}}}_{i}\\
\rho_{\nu}(p,t) & \text{otherwise}
\end{cases}\label{eq:rho evolution}\\
\nu(t+{\rm d}t) & =\nu(t)-(\nu-\nu^{*}){\rm d}t+\sqrt{2\sigma^{2}}{\rm d}W(t)\label{eq:OU differentiel}
\end{align}
The second equation (\ref{eq:OU differentiel}) is just another way
to write Equation (\ref{eq:OU process}). We have ${\rm d}W_{1}={\rm d}W_{2}$,
${\rm d}W_{3}={\rm d}W_{4}$, and $\left\langle {\rm d}W_{1}(t){\rm d}W_{1}(t')\right\rangle =\delta(t-t'){\rm d}t$
and $\left\langle {\rm d}W_{1}(t){\rm d}W_{3}(t')\right\rangle =0$.
The consequence of equation (\ref{eq:rho evolution}) is that at each
step, the probability distribution is constant over each region $\mathcal{R}_{i}$.
But as the reader can see on equation (\ref{eq:rho evolution}), the
region $\mathcal{R}_{i}$ is moving because of the variations of the
frequencies $\left\{ \nu_{i}\right\} _{i=1..4}$. Therefore, at the
next step, the average is performed over a region which has slightly
moved during the time step ${\rm d}t$.\\
\\

\subsection{The Markov process that corresponds to the local diffusion model }

In this section, we give a rigorous mathematical definition of the
process described by equations (\ref{eq:rho evolution}-\ref{eq:OU differentiel}).
We explain that there is no proper Markov process for the variable
$(p,\nu)$ corresponding exactly to equations (\ref{eq:rho evolution}-\ref{eq:OU differentiel})
but that an equivalent Markov process exists on an extended space.
For simplicity, we define the Markov process for a single diffusion
region $\mathcal{R}$, but the following discussion can be straightforwardly
generalized to many diffusion regions. \\

Let us consider the stochastic process $(p_{t},\nu_{t})$ defined
by the following rules. First, $\text{d}\nu_{t}=-(\nu_{t}-\nu^{*}){\rm d}t+\sqrt{2\sigma^{2}}{\rm d}W(t)$,
and then $p_{t}$ is a jump process such that if $p_{t}\in\mathcal{R}$
then $p_{t}$ jumps to any other point of $\mathcal{R}$ at a rate
$1/(\epsilon\delta)$ (where $\delta$ is the width of the diffusion
region), and that if $p_{t}$ does not belong to $\mathcal{R}$ then
it stays constant. This defines a Markov process. One can think of
$\epsilon$ as being the same as in section \ref{subsec:Formulation-of-the},
or being another unrelated parameter. One can write the infinitesimal
generator of this process. If $\phi$ is a test function on the space
$(p,\nu)$, then the infinitesimal generator is
\begin{equation}
G_{(p,\nu)}\left[\phi\right]=\frac{1}{\epsilon}\left[\frac{1}{\delta}\int_{\nu-\delta/2}^{\nu+\delta/2}\phi\left(p_{1},\nu\right)\,\text{d}p_{1}-\phi\left(p,\nu\right)\right]\text{I}_{\mathcal{R}}(p)+L_{\nu}\left[\phi\right],\label{eq:finite-rate-local-diffusion-model}
\end{equation}
where $\text{I}_{\mathcal{R}}(p)$ is the step function equal to 1
if $p\in\mathcal{R}$ and zero otherwise, and $L_{\nu}$ is the infinitesimal
generator of the diffusion (\ref{eq:OU differentiel}).

The process described by (\ref{eq:rho evolution}-\ref{eq:OU differentiel})
corresponds to the infinite rate limit ($\epsilon\rightarrow0$ or
$1/\epsilon\rightarrow\infty$) of the process with infinitesimal
generator (\ref{eq:finite-rate-local-diffusion-model}). We could
equivalently say that (\ref{eq:rho evolution}-\ref{eq:OU differentiel})
is the infinite rate limit of the finite rate jump process (\ref{eq:finite-rate-local-diffusion-model}).
As is clearly seen from (\ref{eq:finite-rate-local-diffusion-model}),
the infinitesimal generator (\ref{eq:finite-rate-local-diffusion-model})
does not have a simple limit when $\epsilon\rightarrow0$. In this
limit, when $p_{t}$ is inside $\mathcal{R}$, it oscillates faster
and faster, uniformly over the set $\mathcal{R}$. In order to define
properly the limit for such a fast oscillating variable, we would
need the formalism of Young measures and weak convergence. 

For finite $\epsilon$, we first define a Markov process over the
space of measures as follows: we first regularize the process $\nu_{t}$
by introducing a correlation time $\tau_{c}$, and $\nu_{t}$ is the
diffusion process
\begin{equation}
\dot{\nu_{t}}=-(\nu_{t}-\nu^{*})+\sqrt{2\sigma^{2}}\eta(t),\label{eq:regularized nu}
\end{equation}
where $\eta$ is a continuous random process such that $\left\langle \eta(t),\eta(t')\right\rangle =\frac{1}{\tau_{c}}e^{-t/\tau_{c}}$.With
the above definition, the velocity $\dot{\nu_{t}}$ is a well-defined
variable. The mixing region is still defined as $\mathcal{R}_{t}:=[\nu_{t}-\delta/2,\nu_{t}+\delta/2]$.
Then we introduce the process $\mu_{t}$ in the space of measures
and the jump process $p_{t}^{0}\in\mathbb{R}$ with the following
rules:
\begin{enumerate}
\item The state $(\mu_{t}=\delta(p-p_{t}^{0}),p_{t}^{0})$ is invariant
if $p_{t}^{0}\notin\mathcal{R}$. This law translates in measure space
the fact that the jump process $p_{t}$ does not move outside $\mathcal{R}_{t}$.
\item The process $(\mu_{t},p_{t}^{0})$ jumps from $(\mu_{t}=\delta(p-p_{t}^{0}),p_{t}^{0})$
to $(\mu_{t}=I_{\mathcal{R}}(p),p_{t}^{0})$ with rate $\frac{1}{\epsilon}$
when $p_{t}^{0}\in\mathcal{R}_{t}$. This law translates in measure
space the fact that the jump process $p_{t}$ can jump to any point
in $\mathcal{R}_{t}$ with rate $1/(\epsilon\delta)$.
\item We define the variable $B\mathcal{R}_{t}:=\nu_{t}-sgn(\dot{\nu}_{t})\frac{\delta}{2}$,
which corresponds to the boundary of $\mathcal{R}_{t}$ opposite to
the moving direction of $\nu_{t}$. The process $(\mu_{t},p_{t}^{0})$
jumps from $(\mu_{t}=I_{\mathcal{R}}(p),p_{t}^{0})$ to $(\mu_{t}=\delta(p-(B\mathcal{R}_{t})),B\mathcal{R}_{t})$
with rate $\frac{\dot{\nu}_{t}}{\delta}$. This corresponds to the
exit from region $\mathcal{R}_{t}$. With the regularization (\ref{eq:regularized nu})
of the process $\nu_{t}$, the jumping rate is perfectly defined.
\end{enumerate}
The Markov process $(\nu_{t},\mu_{t},p_{t}^{0})$ described by the
rules 1,2 and 3 is equivalent to the Markov jump process $p_{t}$
which infinitesimal generator is given by (\ref{eq:finite-rate-local-diffusion-model}).
With the formalism of Young measures, the $\epsilon\rightarrow0$
limit becomes straightforward: the process converges to the jump process
defined by rules 1 and 3, and with the infinite rate limit for the
second rule. This means that the process $(\mu_{t},p_{t}^{0})$ jumps
instantaneously from $(\mu_{t}=\delta(p-p_{t}^{0}),p_{t}^{0})$ to
$(\mu_{t}=I_{\mathcal{R}}(p),p_{t}^{0})$ as soon as $p_{t}^{0}\in\mathcal{R}_{t}$.
This gives a precise definition of the local diffusion model (\ref{eq:rho evolution}-\ref{eq:OU differentiel}).
We note however that the limit $\tau_{c}\rightarrow0$ in this model
leads to some troubles as the jump rate $\frac{\dot{\nu}_{t}}{\delta}$
becomes infinite. The question of the limit $\tau_{c}\rightarrow0$
is a very interesting mathematical one, but it is out of the scope
of this paper.\\

In this subsection, we have set up the mathematical framework. We
have built two models: the first model is a Hamiltonian model (\ref{eq:toy hamiltonian})
that depends on time and on frequencies with a slow stochastic evolution
(\ref{eq:OU process}). The second model is completely stochastic,
and can be thought of as the ``averaging'' of the first model. It
is called the local diffusion model, and defined by equations (\ref{eq:rho evolution}-\ref{eq:OU differentiel}).
The great interest of the local diffusion model is that it is completely
stochastic, and thus much simpler to study than the first model, which
still keeps the complexity inherent to a chaotic dynamics. 

We still did not explain on which conditions the local diffusion model
should give relevant predictions for the stochastic Hamiltonian model,
this is one crucial question that is answered in section \ref{subsec:Numerical-simulations}.
Section \ref{subsec:Numerical-simulations} presents the numerical
simulations performed with the first model (\ref{eq:toy hamiltonian}-\ref{eq:OU process}),
with comparisons with the local diffusion model. We study in section
\ref{subsec:Numerical-simulations} first exit times $\tau$ from
a domain, both for the model (\ref{eq:toy hamiltonian}-\ref{eq:OU process})
and for the local diffusion model. More precisely, we define the variable
$\tau$ as the first time to reach $p=0$ starting from $p=10$. Our
aim is then to compute numerically the probability distribution $\rho(\tau)$.
In particular, we focus on the typical time $\tau^{*}$ where the
probability distribution $\rho(\tau)$ reaches its maximal value,
and on the dependance of the distribution on the noise amplitude $\sigma$.

\subsection{Comparison of the dynamics of the stochastic Hamiltonian model with
the local diffusion model\label{subsec:Numerical-simulations}}

\subsubsection{Simulations of the Hamiltonian dynamics with stochastic frequencies}

We now present the numerical results obtained with the stochastic
Hamiltonian model defined by (\ref{eq:toy hamiltonian}-\ref{eq:OU process}).
The Hamiltonian (\ref{eq:toy hamiltonian}) has the form $A(p)+B(q,t)$
with
\begin{align*}
A(p) & :=\frac{p^{2}}{2},\\
B(q,t) & :=\stackrel[k=1]{4}{\sum}\cos\left(q-\frac{1}{\epsilon}\int_{0}^{t}\nu_{k}(s){\rm d}s-\varphi_{k}\right).
\end{align*}
 We have thus used the symplectic integrator of order 4 $SBAB_{3}$
\cite{yoshida1990construction}. At each time step, we integrate the
frequencies from equation (\ref{eq:OU process}) with a stochastic
Euler algorithm. The parameter $\epsilon$, the mean frequencies $\nu_{k}^{*}$
and the initial phases $\varphi_{k}$ were fixed to their nominal
values given in Table (\ref{tab:parameters}). 

As we have explained at the end of Section \ref{subsec:averaging-of-the},
we are mainly interested in the first exit time $\tau$ defined as
the first time to leave the region $p>0$ starting from $p=10$. We
want to compute numerically its probability distribution function
$\rho(\tau)$, and determine how it depends on the noise amplitude
$\sigma$. To achieve this aim, we have performed a set of five numerical
simulations using the values of $\sigma$ given in Table \ref{tab:sigma values}.
For each simulation, we ran 5000 trajectories all starting at the
same point $(p,q)=(10,0)$. Each trajectory is run with a different
realization of the noise $W(t)$. The results of the simulations are
the histograms displayed in figure (\ref{fig:results}). The histograms
represent the distributions $\rho(\tau)$ for each simulation.

\subsubsection{Simulations of the local diffusion model}

\begin{table}
\begin{centering}
\begin{tabular}{|c|c|c|}
\hline 
Simulation & $\sigma$ & $T_{max}$ $\times\frac{\nu_{1}}{2\pi}.10^{-4}$\tabularnewline
\hline 
\hline 
1 & 3.0 & 1.0\tabularnewline
\hline 
2 & 2.2 & 2.0\tabularnewline
\hline 
3 & 1.84 & 3.0\tabularnewline
\hline 
4 & 1.1 & 10.0\tabularnewline
\hline 
5 & 0.7 & 50.0\tabularnewline
\hline 
\end{tabular}
\par\end{centering}
\caption{Noise amplitudes and integration times of the five simulations in
figure (\ref{fig:results}). \label{tab:sigma values}}
\end{table}
The local diffusion model is given by equations (\ref{eq:rho evolution}-\ref{eq:OU differentiel}).
So far, we did not prescribe the values of the parameters $\delta_{1},\delta_{2}$
corresponding to the extensions of the diffusion patches $\mathcal{R}_{1},\mathcal{R}_{2},\mathcal{R}_{3}$.
$\delta_{1}$ and $\delta_{2}$ should correspond to the effective
extension of the strongly chaotic regions of the stochastic Hamiltonian
model. The parameters $\delta_{1}$ and $\delta_{2}$ could be estimated
from the Chirikov criterion of resonance overlap. However, it is known \cite{lichtenberg2013regular} that the size of the chaotic regions is
smaller than the theoretical predictions of the Chirikov criterion. This is confirmed here by the numerical simulations.
To obtain a better agreement with the simulations, we prescribed the
size of the diffusion patches with the following method.

\begin{figure}
\begin{centering}
\includegraphics[height=8cm]{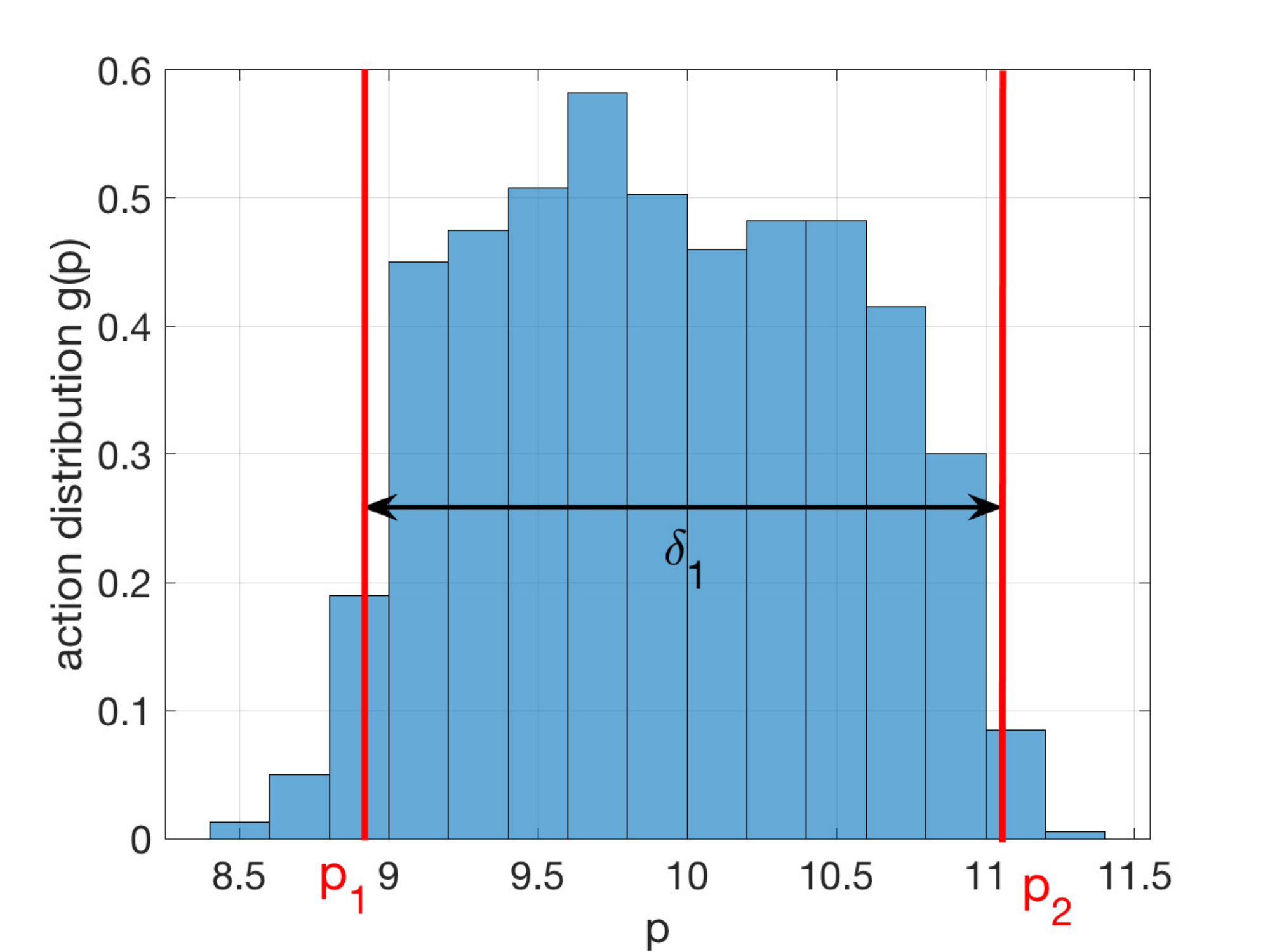}
\par\end{centering}
\caption{Histogram of the density in the chaotic region 1 after about 300 turnover
times. The two red lines at $p_{1}$ and $p_{2}$ show the boundaries
of the strongly chaotic region. The mixing region used in the local
diffusion model is defined as the symmetric interval $[p_{1},p_{2}]$
centered at $p=\frac{\nu_{1}^{*}+\nu_{2}^{*}}{2}$ in which 90\% of
the probability distribution is concentrated. Its extension is given
by $\delta_{1}=p_{2}-p_{1}$. \label{fig:hist region1}}
\end{figure}

To estimate the size of the chaotic region 1 (see figure \ref{fig:schema_model}),
we ran a numerical simulation of Hamilton equations with the Hamiltonian
(\ref{eq:toy hamiltonian}), except that we kept the frequencies fixed
to their reference values $\nu_{i}^{*}$. We simulated 2000 trajectories
with initial conditions $p=\nu_{1}^{*}$ and $q$ equally distributed
over the range $[0,2\pi]$, over a time $T=300*\frac{2\pi}{\nu_{1}^{*}}$.
The final coordinates are then distributed over the chaotic region
1, and only very few of them have exited region 1. Figure (\ref{fig:hist region1})
shows a typical histogram of final momenta. We then define the boundaries
of the chaotic region as the symmetric interval $[p_{1},p_{2}]$ centered
at $p=\frac{\nu_{1}^{*}+\nu_{2}^{*}}{2}$ in which 90\% of the probability
distribution is concentrated. Then, the empirical estimate of $\delta_{1}$
is $\delta_{1}\approx p_{2}-p_{1}$.

Using the same method, we have estimated the extension $\delta_{2},\delta_{3},\delta_{4}$
of the chaotic regions corresponding to the resonances of second,
third and fourth orders respectively. The regions are located around
the values $p=3.3$, $p=6.6$ for the resonances of third order, and
$p=2.5$ and $p=7.5$ for the resonances of fourth order. The values
of $\left\{ \delta_{i}\right\} _{i=1..4}$ are reported in table (\ref{tab:delta table}).

We ran $M=1000$ simulations using for each a new realization of the
stochastic process $\nu(t)$. For each of the realization $\nu(t)$,
we could compute with (\ref{eq:rho evolution}) the density $g_{\nu}(p,t)$
of the probability to find the system at $p$ at time $t$. At the
beginning, the system is in $p=\nu_{1}^{*}$, which corresponds to
the initial condition $g_{\nu}(p,0)=\delta(p-\nu_{1}^{*})$. We want
to compute the probability of first hitting time at $p=0$. This means
that we have to prescribe the boundary condition $g_{\nu}(p<0,t)=0$.
In practice, this condition amounts to set $g_{\nu}(p,t)=0$ for $p\in[\nu_{4}(t)-\delta_{1}/2,\nu_{4}(t)+\delta_{1}/2]$
because if the system enters in the diffusion patch $\mathcal{R}_{2}$,
it is immediately transported across the patch and reaches $p=0$. 

The complete probability distribution $g(p,t)$ is simply the average
of $g_{\nu}(p,t)$ over the realizations of the stochastic process
$\nu(t)$. Let $\left\{ \nu^{k}(t)\right\} _{k=1..M}$ be the $M$
realizations of $\nu(t)$, we have 
\[
g(p,t)=\frac{1}{M}\stackrel[k=1]{M}{\sum}g_{\nu^{k}}(p,t).
\]
Then the probability of first hitting times $g(\tau)$ is given by
\[
\rho(\tau)=-\frac{{\rm d}}{{\rm d}\tau}\int_{0}^{+\infty}g(p,\tau){\rm d}p.
\]

The simulations are performed with the set of parameters given in
table (\ref{tab:sigma values}). The results are displayed on the
different graphs of figure (\ref{fig:results}) together with the
histograms obtained by the direct Hamiltonian simulations. The curves
show the results of $\rho(\tau)$ obtained with the simulations of
the local diffusion model, whereas the histograms show the results
for $\rho(\tau)$ obtained with the simulations of the stochastic
Hamiltonian model. On the simulations 1, 2 and 3, we have only used
the local diffusion model including the resonances up to second order.
But on the simulations 4 and 5, there are two curves for $\rho(\tau)$:
the lower one is the distribution $\rho(\tau)$ computed with the
local diffusion model with resonances up to second order, but on simulation
4 and 5, we included the resonances up to fourth order in the local
diffusion model. The results are displayed by the upper curves in
simulations 4 and 5.

\begin{table}
\begin{centering}
\begin{tabular}{|c|c|c|}
\hline 
resonances & extension of the chaotic region & numerical estimation of $p_{2}-p_{1}$\tabularnewline
\hline 
\hline 
first order  & $\delta_{1}$ & 2.25\tabularnewline
\hline 
second order & $\delta_{2}$ & 0.50\tabularnewline
\hline 
third order & $\delta_{3}$ & 0.28\tabularnewline
\hline 
fourth order & $\delta_{4}$ & 0.09\tabularnewline
\hline 
\end{tabular}
\par\end{centering}
\caption{Width of the diffusion patches of the local diffusion model. \label{tab:delta table}}
\end{table}

\subsection{Discussion}

\begin{figure}
\begin{centering}
\includegraphics[height=4.5cm]{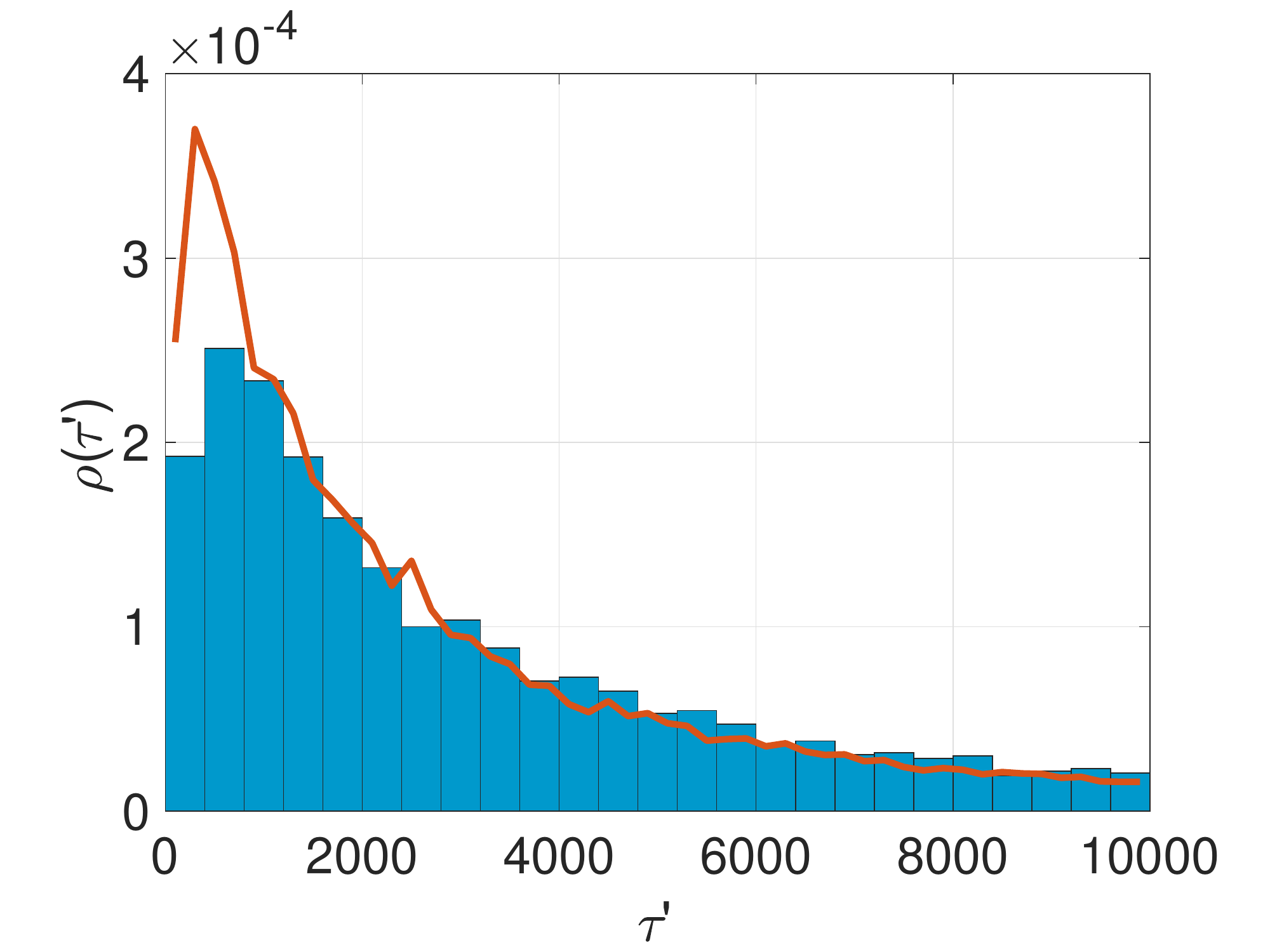}\includegraphics[height=4.5cm]{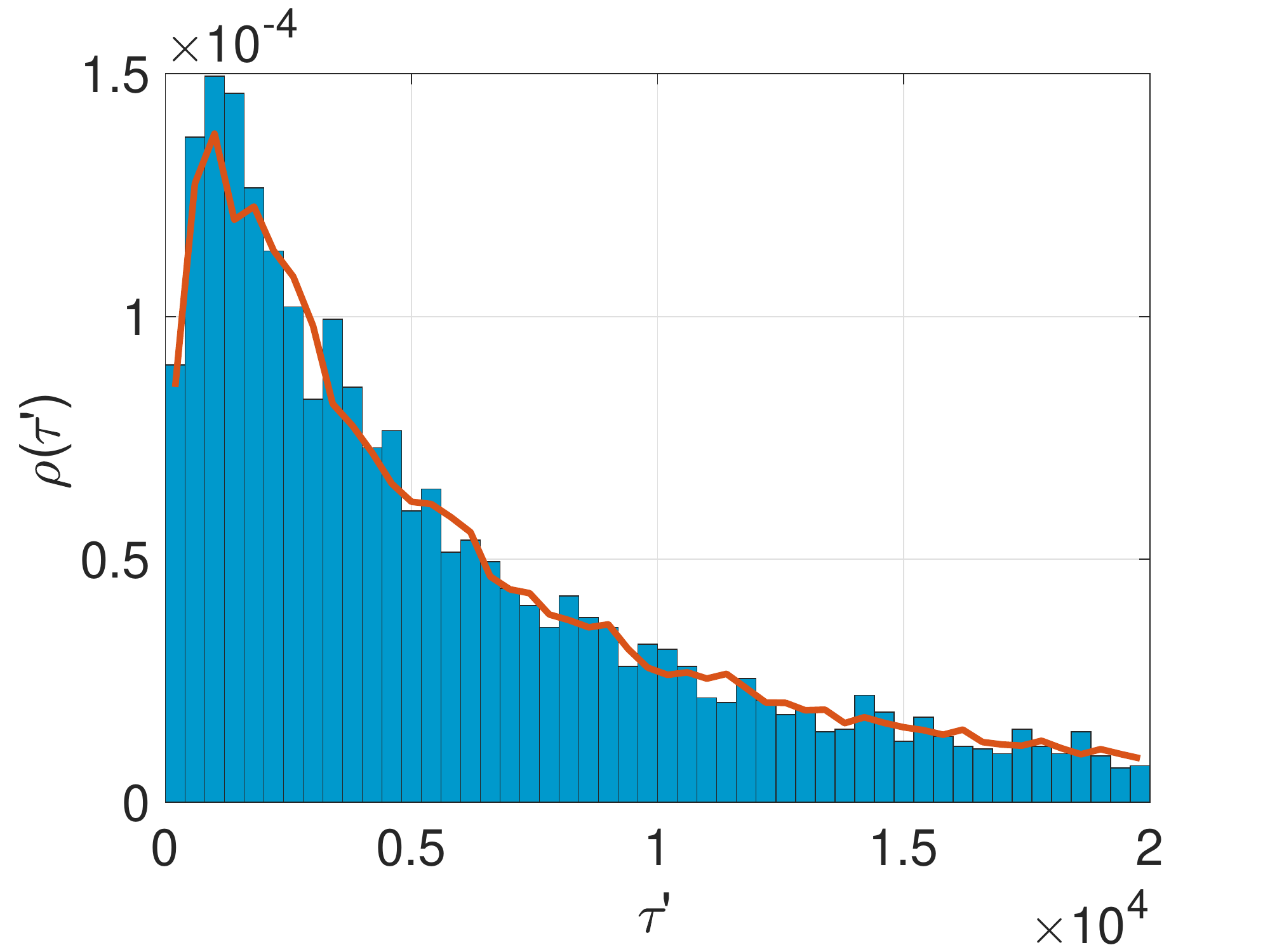}\includegraphics[height=4.5cm]{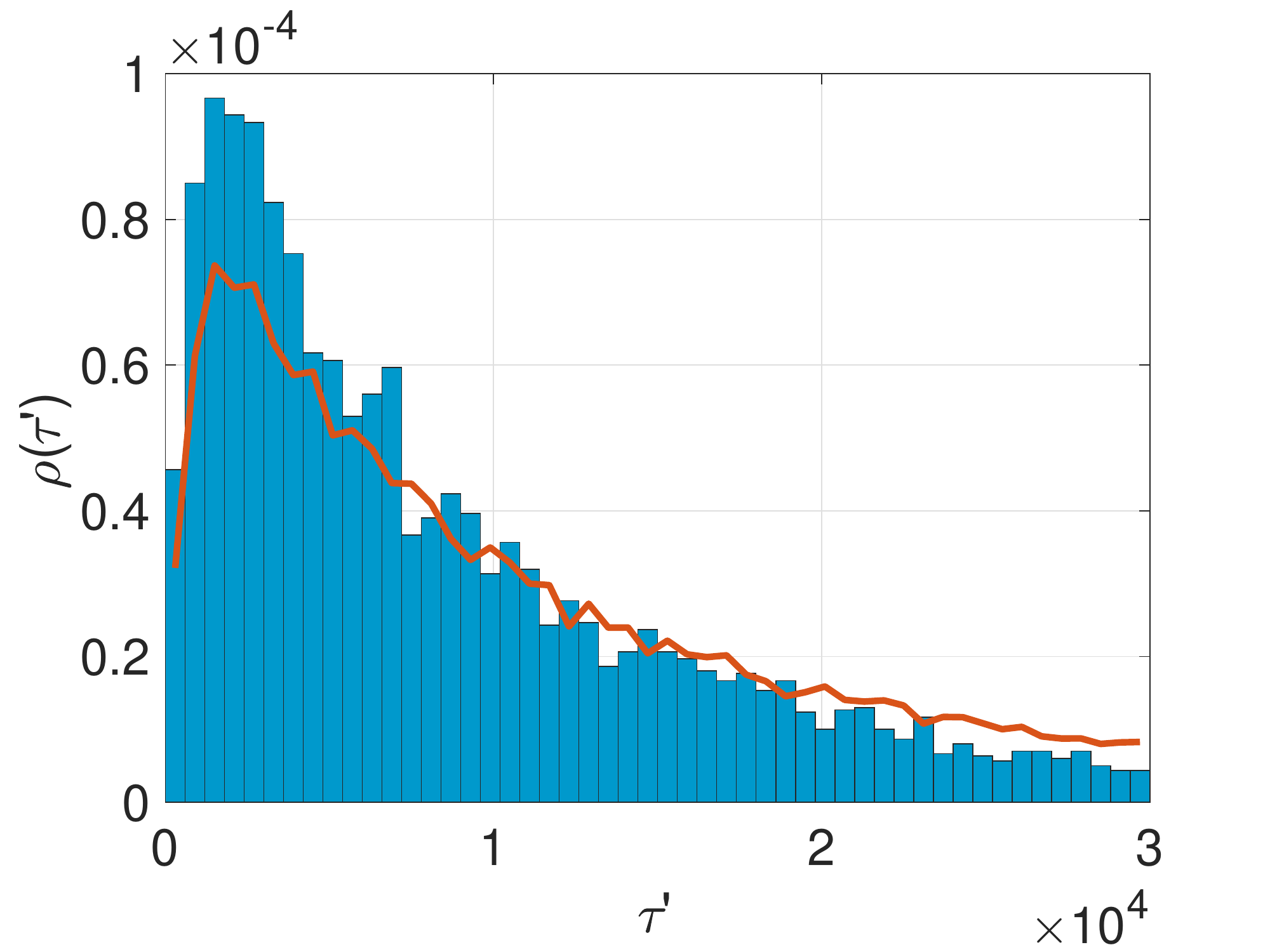}
\par\end{centering}
\caption{First exit time distributions $\rho(\tau)$ for three simulations
with $\sigma=3.0$ (left), $\sigma=2.2$ (middle) and $\sigma=1.84$
(right). We display the distributions in terms of the non-dimensional
time $\tau':=\tau\times\frac{\nu_{1}}{2\pi}$. The histograms display
the results of the simulations with the Hamiltonian (\ref{eq:toy hamiltonian}).
The curves show the results of the corresponding local diffusion model
including resonances up to second order. \label{fig:results}}
\end{figure}

The different simulations in figure (\ref{fig:results}) aim at showing
which phenomenon governs the transport in phase space in the stochastic
Hamiltonian model (\ref{eq:toy hamiltonian}-\ref{eq:OU process}),
and how the transport depends on the parameters of the model. 

One trivial but crucial conclusion of our numerical study is that
the transport in a stochastic Hamiltonian model is very different
from the transport with the same Hamiltonian (\ref{eq:toy hamiltonian})
without stochastic variations of the frequencies. If the frequencies
are fixed, the trajectories starting at $p=\nu_{1}^{*}$ are just
spread across the first mixing region, and none of them reaches the
value $p=0$ within the time $T_{max}$. Transport with stochastic
frequencies is thus a new mechanism that completely overcomes chaotic
diffusion in deterministic chaotic Hamiltonian dynamics.

The qualitative shape of the distributions of first hitting times
$\rho(\tau)$ displayed in figure (\ref{fig:results}) is typical
of a distribution of first exit times from a domain in a stochastic
system. The probability distribution has a maximum $\rho^{*}$ reached
at $\tau^{*}$, that can be considered as the typical time for the
exit event to occur. For times smaller than $\tau^{*}$, the probability
distribution goes rapidly to zero. It is thus very rare for the system
to reach $p=0$ in a time much smaller than the typical time $\tau^{*}$,
because it corresponds to exceptionally large and fast random fluctuations
of the stochastic frequencies $\left\{ \nu_{i}\right\} $. For times
larger than the typical time $\tau^{*}$, the probability distribution
also goes to zero because it is also a rare event, called ``persistence'',
that the system does not leave the domain $p>0$ for a long time. 

The order of magnitude of $\tau^{*}$ depends on the noise amplitude
$\sigma$ acting on the frequencies. For $\sigma=3.0$, it is of the
order of $10^{3}*\frac{2\pi}{\nu_{1^{*}}}$, whereas for $\sigma=0.7$,
it is two orders of magnitude larger, of the order of $10^{5}*\frac{2\pi}{\nu_{1}^{*}}$.
We have checked numerically that the typical exit time $\tau^{*}$
does not depend on $\epsilon$ in the limit $\epsilon\rightarrow0$.
This fact shows that the transport mechanism has a well defined limit
when $\epsilon\rightarrow0$, and confirm the relevance of the local
diffusion model.\\

\begin{figure}
\begin{centering}
\includegraphics[height=5cm]{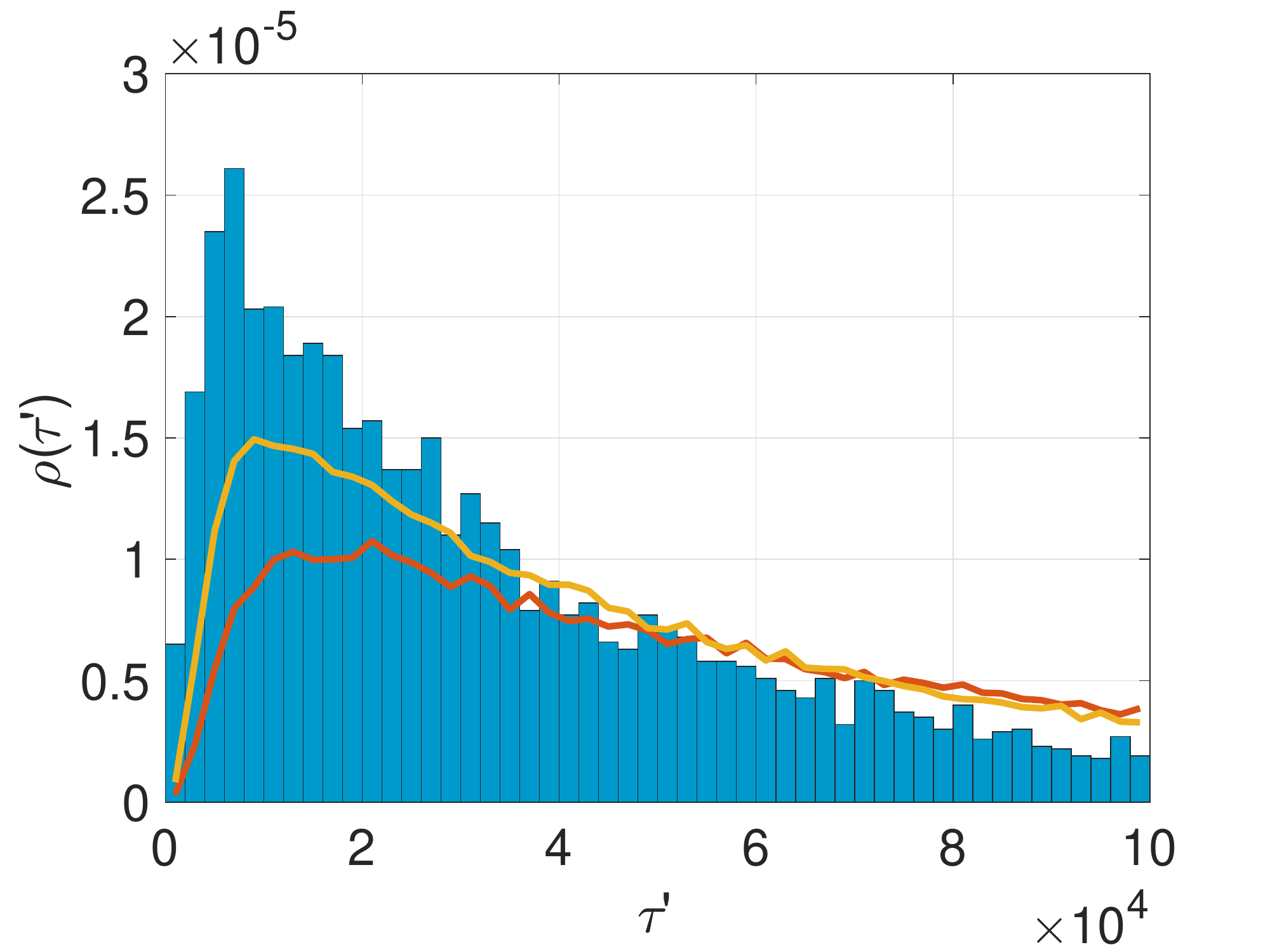}\includegraphics[height=5cm]{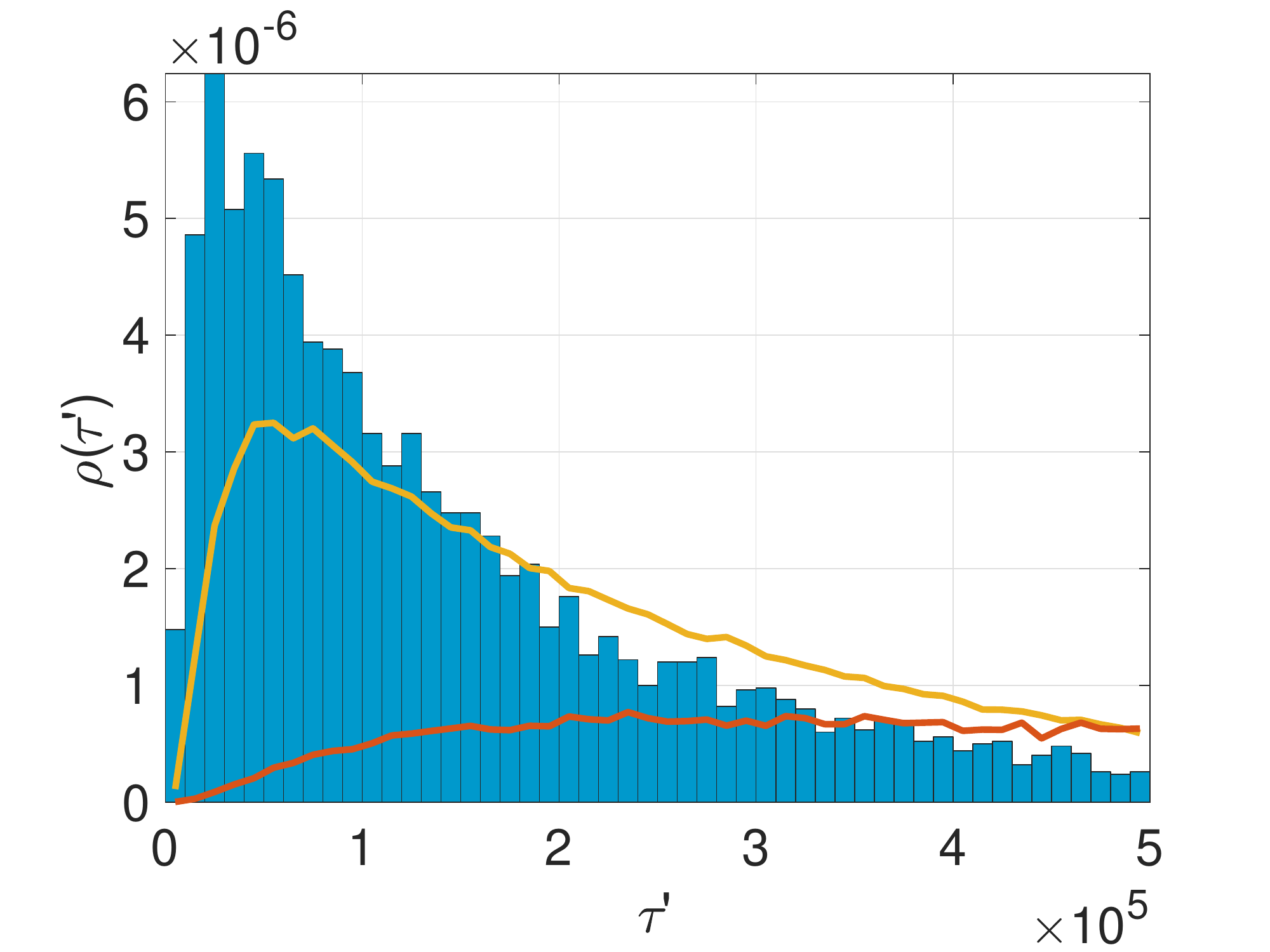}
\par\end{centering}
\caption{First exit time distributions $\rho(\tau)$ for two simulations with
$\sigma=1.1$ (left) and $\sigma=0.7$ (right). We display the distributions
in terms of the non-dimensional time $\tau':=\tau\times\frac{\nu_{1}}{2\pi}$.
The histograms display the results of the simulations with the Hamiltonian
(\ref{eq:toy hamiltonian}). The lower curves (red) show the results
of the corresponding local diffusion model including resonances up
to second order, and the upper curves (yellow) show the results of
the local diffusion model including resonances up to fourth order.\label{fig:results2}}
\end{figure}
The local diffusion model presented in Section \ref{subsec:averaging-of-the}
can be seen as the averaged dynamics of the stochastic Hamiltonian
model, for which transport outside of the principal chaotic regions
has been neglected. This is reflected in the local diffusion model
by the fact that the diffusion coefficient is zero outside the diffusion
patches $\mathcal{R}_{i}$. Therefore, the local diffusion model only
takes into account transport by migration with the chaotic regions.
If the transport of this type is dominant, it is natural to expect
that the local diffusion model reproduces well the results of the
stochastic Hamiltonian model. If, on contrary, transport is mainly
due to the mechanism of noise driven transport in regular regions,
then the exit rate at $p=0$ predicted by the local diffusion model
is much slower than the real transport of the Hamiltonian dynamics
with stochastic parameters. 

On figure (\ref{fig:results}), it can be seen that the local diffusion
model is able to capture quite well the probability distribution of
first exit times $\rho(\tau)$. For the three simulations with $\sigma=3.0/2.2/1.84$,
the local diffusion model including resonances up to second order
gives excellent results. It reproduce qualitatively and quantitatively
the histogram of $\rho(\tau)$, with the same location $\tau^{*}$
of the maximal value, and reproduces the decrease of the distribution
$\rho(\tau)$ for long times. For the two simulations with $\sigma=1.1/0.7$
the local diffusion model of second order predicts a transport rate
which is much smaller than the real transport. In particular for $\sigma=0.7$,
even the qualitative shape of $\rho(\tau)$ for the Hamiltonian model
is not reproduced, the typical time $\tau^{*}$ is by far overestimated.
This means that for $\sigma$ values below 1.84, transport through
resonances of order higher than two can no longer be neglected. The
local diffusion model including resonances up to fourth order is able
to reproduce qualitatively the distribution of exit times $\rho(\tau)$,
and gives also a good order of magnitude of the value of $\rho(\tau)$.
This is illustrated by the upper curves for the simulations 4 and
5 on figure (\ref{fig:results}). For values of $\sigma$ below $0.7$,
the local diffusion model is no longer able to reproduce transport
in phase space, even with resonances up to fourth order. This comes
from the fact that the overlap of two neighboring chaotic regions
becomes very rare when $\sigma$ decreases, and thus the mechanism
of noise driven transport in the regular regions overcomes the first
mechanism. \\

The relevance of the local diffusion model to predict transport depends
on the balance between the amplitude $\sigma$ of frequencies fluctuations
and the distance to cross between two neighboring diffusion regions.
For example, in the local diffusion model of order two, the initial
distance between two neighboring diffusion patches is 
\[
\text{Dist}_{\mathcal{R}_{1}\rightarrow\mathcal{R}_{3}}:=\left(\frac{\nu_{1}^{*}+\nu_{2}^{*}}{2}-\frac{\delta_{1}}{2}\right)-\left(\frac{\nu_{1}^{*}+\nu_{4}^{*}}{2}+\frac{\delta_{2}}{2}\right)\approx3.575.
\]
The sum of the variances of the fluctuations of the diffusion patches
$\mathcal{R}_{1}$ and $\mathcal{R}_{3}$ gives the typical amplitude
of the fluctuations of the mixing regions 
\[
\sqrt{V_{\mathcal{R}_{1}}+V_{\mathcal{R}_{3}}}=\left(1+\frac{1}{\sqrt{2}}\right)\sigma,
\]
where $V_{\mathcal{R}_{i}}$ is the variance of fluctuations of region
$i$. Therefore, the efficiency of the transport through migration
of the mixing regions depends on whether the parameter 
\[
\Delta:=\frac{\text{Dist}_{\mathcal{R}_{1}\rightarrow\mathcal{R}_{3}}}{\sqrt{V_{\mathcal{R}_{1}}+V_{\mathcal{R}_{3}}}}\simeq\frac{2.1}{\sigma}
\]
is large or small compared to one.

This means that for the three simulations with $\sigma=3.0/2.2/1.84$,
the initial distance to cross between two diffusion patches is larger
or of same order as the amplitude of the frequency fluctuations. The
jump between one diffusion patch to another is thus possible with
``typical'' fluctuations of the frequencies. Transport does not
require an exceptionally large fluctuation. But this is no longer
the case for $\sigma=1.1/0.7$. Frequency fluctuations are two small
to pass directly from the mixing region surrounding the first order
resonances to the mixing region surrounding the second order resonances
(or equivalently, from $\mathcal{R}_{1}$ to $\mathcal{R}_{3}$),
and transport is due to higher order resonances. For example, in the
local diffusion model of fourth order, the initial distance to cross
to jump from the diffusion patch $\mathcal{R}_{1}$ to the next patch
$\mathcal{R}_{.}$ is $\text{Dist}_{\mathcal{R}_{1}\rightarrow\mathcal{R}_{.}}\left(\frac{\nu_{1}^{*}+\nu_{2}^{*}}{2}-\frac{\delta_{1}}{2}\right)-\left(\frac{3\nu_{1}^{*}+\nu_{4}^{*}}{2}+\frac{\delta_{4}}{2}\right)\approx1.28$.
On the other hand, the combined fluctuations of the two patches have
a variance of $\sqrt{V_{\mathcal{R}_{1}}+V_{\mathcal{R}_{.}}}\left(1+\frac{\sqrt{10}}{4}\right)\sigma$.
Thus the parameter $\Delta$ is of the order of $\Delta\simeq\frac{1.28}{1+\sqrt{10}/4}\frac{1}{\sigma}\simeq\frac{0.71}{\sigma}$.
This argument explains why the local diffusion model up to order four
is able to predict the transport for values of $\sigma$ of the order
of $0.7$, but fails for lower values of $\sigma$.\\

In the present section, we have shown how a system that satisfies
a Hamiltonian dynamics with stochastic frequencies can be transported
slowly through phase space by the slow displacement of the chaotic
regions. We have shown that this kind of transport can be reproduced
qualitatively and quantitatively by a Markov model, the local diffusion
model. The local diffusion model gives a representation of the strongly
chaotic regions created by resonance overlap by diffusion patches
with infinite diffusion coefficient. The relevance of this Markov
model to predict transport rates in the stochastic Hamiltonian model
mainly depends on the amplitude of the frequency stochastic variations.
We have shown that for decreasing amplitudes of the variations, the
local diffusion model should take into account resonances of higher
and higher orders. In this section, we have presented a model including
resonances up to order four. But one cannot expect the local diffusion
model to be valid for all range of the amplitude fluctuations $\sigma$,
even if we include resonances up to higher orders. The reason for
that is that transport is also due to noise driven transport in the
regular regions, away from the resonances. If the amplitude of the
fluctuations is too small, then transport is mainly due to noise driven
transport in the regular regions. We have shown that transport is
mainly due to migration with the chaotic regions if the typical fluctuations
of the frequencies are similar to the distance in phase space between
two neighboring chaotic regions. 

We have performed an other numerical simulation where the stochastic
process for the variations of $\nu$ is a jump process with exponential
distribution of the jumps, and we found that the results are in agreement
with the general picture we give in this section. This suggests that
the transport mechanism described in this article is robust to a change
in the stochastic process for the evolution of the external parameter.

\section{Conclusion}

We have studied time dependent Hamiltonian systems with one degree
of freedom, for which the Hamiltonian depends on an additional slow
stochastic parameter. The slowly varying parameter introduces a timescale
separation in the system, which allows us to use the theory of averaging
to describe the long term evolution of the system. When the fast Hamiltonian
dynamics is integrable, it has been shown that the slow evolution
can be described by a diffusion process of the action variable. This
first part of our work is mostly an extension of the theory of adiabatic
invariants to the stochastic case. Because  of the irregularity of the stochastic process, adiabatic invariants are not conserved in the limit of a large timescale separation, unlike in the classical theory. This gives a new mechanism of transport that we call "noise driven transport in regular regions".

More interesting for practical
applications is the case where the fast Hamiltonian dynamics is chaotic.
We have shown that transport in phase space comes from the slow displacement
of chaotic regions, and is equivalent, for some ranges of the parameters,
to a Markov process called the local diffusion model. We have shown
numerically that the local diffusion model give quantitative results
in agreement with the full simulations of the Hamiltonian dynamics
with stochastic parameters. 

This work can be seen as a pioneer work on a new class of dynamical systems, Hamiltonian dynamical systems with slowly changing phase space structure. By no mean does this work intend to be exhaustive on the subject. In particular, our work leaves many open questions. First, it has been shown it section \ref{sec:The-integrable-case} that the mechanism of noise-driven transport comes from the strong irregularity of the stochastic process. One could wonder what  happens when the stochastic process is smooth, with a finite correlation time. How does transport depend on this new timescale? And how does the transition happens to the white noise limit presented in our work? Also, while our theory do not consider separatrix crossing for integrable dynamics, it is known that separatrix crossing occurs in numerous physical situations and an extension of our theory would be interesting in this case. In section \ref{sec:The-chaotic-case}, we focused on the mechanism of "transport by slow migration with the chaotic regions". A full description of transport in the chaotic case presented in section \ref{sec:The-chaotic-case} should also include extension of the chaotic regions, that was neglected in the present work.

The reduction of the dynamics to a Markov process opens the possibility
to use large deviation theory to compute the probability of rare events
in the dynamical system, for example the probability of an exceptionally
fast exit out of a domain. Simplified Hamiltonian models with few
degrees of freedom and a stochastic parameter can be used as preliminary
works to find qualitative behaviors and order of magnitudes before
resorting to involved numerical simulations. It is hoped that our
work will find interesting applications in celestial mechanics and
other domains of physics.

\section*{Acknowledgement}

We thank Cristel Chandre for his help in the preliminary stage of
this work. We also thank Jacques Laskar whose work on the chaotic
obliquity of planets has been the main motivation of the present work.
The research leading to these results has received funding from the
European Research Council under the European Union\textquoteright s
seventh Framework Program (FP7/2007-2013 Grant Agreement No. 616811).
\\

\bibliographystyle{unsrt}
\bibliography{biblio_dynamics}

\appendix

\section{Averaging in slow-fast system when the slow dynamics is a diffusion
process\label{sec:averaging}}

In the present section, we explain how to average the slow-fast system
(\ref{eq:good slow-fast}) to obtain closed stochastic differential
equations for the variables $\nu$ and $P$. The difficult step comes
from the average of the stochastic term
\[
{\rm d}\xi:=\sigma(Q,P,\nu){\rm d}W,
\]
 where $\sigma$ is the vector
\[
\sigma:=\begin{pmatrix}b(\nu)\\
b(\nu)\frac{\partial H_{1}}{\partial Q}(Q,P,\nu)
\end{pmatrix},
\]
 and $W$ is a one-dimensional Wiener process. 

A variation $\Delta\xi$ over a given fixed time interval $\Delta t$
is given by
\begin{equation}
\Delta\xi=\int_{0}^{\Delta t}\sigma(Q(t'),P(t'),\nu(t')){\rm d}W(t').\label{eq:variation stoch}
\end{equation}
According to the first equation of (\ref{eq:good slow-fast}), $Q$
is a fast variable, which means that it is in fact a function of $\frac{t}{\epsilon}$,
with $\epsilon\ll1$. If we consider a time interval $\Delta t$ small
enough, the increment of the functions $\nu(t)$ and $P(t)$ during
$\Delta t$ can be considered as constant in (\ref{eq:variation stoch})
to leading order. This assumption will turn out to be self-consistent.
Accordingly, we can write (\ref{eq:variation stoch}) as
\begin{equation}
\Delta\xi=\int_{0}^{\Delta t}\sigma(Q(t'/\epsilon),P,\nu){\rm d}W(t').\label{eq:variation stoch-1}
\end{equation}
Then, using the property of self-similarity for the Wiener process,
namely that $W(\epsilon t')=\sqrt{\epsilon}W(t')$, the increment
(\ref{eq:variation stoch-1}) becomes
\begin{equation}
\Delta\xi=\sqrt{\epsilon}\int_{0}^{\Delta t/\epsilon}\sigma(Q(t'),P,\nu){\rm d}W(t').\label{eq:variation stoch-1-1}
\end{equation}
 The stochastic integral in (\ref{eq:variation stoch-1-1}) has to
be understood as a sum of independent Gaussian random variables (see
\cite{gardiner1985stochastic}). Partitioning the interval $\Delta t/\epsilon$
in $N$ small intervals $[t_{i},t_{i+1}]$, the stochastic integral
(\ref{eq:variation stoch-1-1}) is the large $N$ limit of the sum
\begin{equation}
\Delta\xi=\underset{N\rightarrow\infty}{\lim}\sqrt{\epsilon}\stackrel[i=1]{N}{\sum}\sigma(Q(t_{i}),P,\nu){\rm d}W_{i},\label{eq:sum stoch}
\end{equation}
where all the variables ${\rm d}W_{i}$ are independent Gaussian random
variables with zero mean and variance $\frac{\Delta t}{\epsilon N}$.
Now, the key property is that a sum of independent Gaussian random
variables is a random variable. To find the limit in (\ref{eq:sum stoch}),
one just has to compute the variance of the sum. We get
\begin{align*}
\left\langle \left(\Delta\xi\right)^{2}\right\rangle  & =\underset{N\rightarrow\infty}{\lim}\epsilon\stackrel[i,j=1]{N}{\sum}\sigma(Q(t_{i}),P,\nu)\sigma^{T}(Q(t_{j}),P,\nu)\left\langle {\rm d}W_{i}{\rm d}W_{j}\right\rangle ,\\
 & =\underset{N\rightarrow\infty}{\lim}\frac{\Delta t}{N}\stackrel[i=1]{N}{\sum}\sigma\sigma^{T}(Q(t_{i}),P,\nu),
\end{align*}
where the last equality has been obtained using that $\left\langle {\rm d}W_{i}{\rm d}W_{i}\right\rangle =\frac{\Delta t}{\epsilon N}$
, and $\left\langle {\rm d}W_{i}{\rm d}W_{i}\right\rangle =0$ if
$i\neq j$. The final step is to notice that the average
\[
\frac{1}{N}\stackrel[i=1]{N}{\sum}\sigma\sigma^{T}(Q(t_{i}),P,\nu)
\]
is done over the time interval $\frac{\Delta t}{\epsilon}$, with
$\epsilon\ll1$. Over this timescale, the time average corresponds
to the average over the invariant measure of the process $Q(t)$.
The final result is 
\[
\left\langle \left(\Delta\xi\right)^{2}\right\rangle =\Delta t\left\langle \sigma\sigma^{T}(Q,P,\nu)\right\rangle _{Q}.
\]
This justifies the computation of the average in (\ref{eq:average correlations}),
and the result in (\ref{eq:average action}).

It is important to bear in mind that this result is obtained from
the property that the white noise has a vanishing correlation time.
This hypothesis is implicit in the change of timescale in (\ref{eq:variation stoch-1-1}).
The above result would not be true if we had considered, instead of
a diffusion process with white noise, a fast process with a finite
correlation time $\tau_{c}$ of the order of $\epsilon$. The
mechanism of \textquotedbl noise driven transport in regular regions\textquotedbl{}
that we describe in section \ref{sec:The-integrable-case} would not
be the same if the noise in the equation for $\nu$ would have a correlation
time of the same order as the typical time for the variations of $Q$.

\section{Canonical change of variables and existence of a function $H_{1}$
such that $\frac{\partial P}{\partial\nu}=-\frac{\partial H_{1}}{\partial Q}$
and $\frac{\partial Q}{\partial\nu}=\frac{\partial H_{1}}{\partial P}$
\label{sec:Canonical-change-of}}

Let $(p,q)$ be the canonical variables of the Hamiltonian $H(p,q,\nu)$
where is $\nu$ an external parameter of the Hamiltonian. We assume
that for a given value of $\nu$, the Hamiltonian $H$ is integrable,
which means that for any $\nu$ there exist canonical variables $(P(p,q,\nu),Q(p,q,\nu))$
such that $H(p,q,\nu)=\widetilde{H}(P,\nu)$, and $\widetilde{H}$
does not depend on $Q$. By definition, a change of variable is canonical
one whenever the differential two form is conserved: ${\rm d}P\wedge{\rm d}Q={\rm d}p\wedge{\rm d}q$,
or equivalently the Poisson bracket $\{P,Q\}_{p,q}=1$, where $\{f,g\}_{p,q}=\frac{\partial f}{\partial p}\frac{\partial g}{\partial q}-\frac{\partial f}{\partial q}\frac{\partial g}{\partial p}$.
The aim of this section is to prove that there exists a function $H_{1}(P,Q,\nu)$
such that $\frac{\partial P}{\partial\nu}(p,q,\nu)=-\frac{\partial H_{1}}{\partial Q}(Q(p,q,\nu),P(p,q,\nu),\nu)$
and $\frac{\partial Q}{\partial\nu}(p,q,\nu)=\frac{\partial H_{1}}{\partial P}(Q(p,q,\nu),P(p,q,\nu),\nu)$
(relation (\ref{eq:canonical diff}) in the main text).

In order to prove this result, we extend the phase space by introducing
$V$ the canonical momentum associated to $\nu$. We consider
an extended Hamiltonian with two degrees of freedom
\[
H'(p,q,V,\nu):=H(p,q,\nu)+V.
\]
This amounts to saying that $\nu$ has a dynamics, with $\dot{\nu}=\frac{\partial H'}{\partial V}=1$.
We will show that we can find a new momentum $V'$ such that
the change of variables $(p,q,V,\nu)\rightarrow(P,Q,V',\nu')$
is canonical, where $(P(p,q,\nu),Q(p,q,\nu))$ are the canonical variable
defined above for any fixed $\nu$, and the new variable $\nu'$ is
such that $\nu':=\nu$. A change of variable is canonical if and only
if the differential two-form is conserved, which can be expressed as
\begin{equation}
{\rm d}P\wedge{\rm d}Q+{\rm d}V'\wedge{\rm d}\nu'={\rm d}p\wedge{\rm d}q+{\rm d}V\wedge{\rm d}\nu.\label{eq:diff1}
\end{equation}
Then we express explicitly the differentials ${\rm d}P$ and ${\rm d}Q$
as 
\begin{align*}
{\rm d}P & =\frac{\partial P}{\partial p}{\rm d}p+\frac{\partial P}{\partial q}{\rm d}q+\frac{\partial P}{\partial\nu}{\rm d}\nu,\\
{\rm d}Q & =\frac{\partial Q}{\partial p}{\rm d}p+\frac{\partial Q}{\partial q}{\rm d}q+\frac{\partial Q}{\partial\nu}{\rm d}\nu,
\end{align*}
and we use these relations in (\ref{eq:diff1}) to obtain
\begin{equation}
\left\{ P,Q\right\} _{p,q}{\rm d}p\wedge{\rm d}q+\left\{ P,Q\right\} _{p,\nu}{\rm d}p\land{\rm d}\nu+\left\{ P,Q\right\} _{q,\nu}{\rm d}q\land{\rm d}\nu+{\rm d}V'\wedge{\rm d}\nu'={\rm d}p\wedge{\rm d}q+{\rm d}V\wedge{\rm d}\nu.\label{eq:diff2}
\end{equation}
In the first wedge product, the Poisson bracket satisfies $\{P,Q\}_{p,q}=1$,
thus (\ref{eq:diff2}) becomes
\begin{equation}
\left\{ P,Q\right\} _{p,\nu}{\rm d}p\land{\rm d}\nu+\left\{ P,Q\right\} _{q,\nu}{\rm d}q\land{\rm d}\nu+{\rm d}V'\wedge{\rm d}\nu'={\rm d}V\wedge{\rm d}\nu.\label{eq:diff3}
\end{equation}
We now use that $\nu'=\nu$. We look for a function $H_{1}(Q,P,\nu$)
such that the canonical momentum $V'$ has the form $V'=V-H_{1}$.
Using the latter transformation, the equality (\ref{eq:diff3}) becomes
\[
\left(\left\{ P,Q\right\} _{p,\nu}{\rm d}p+\left\{ P,Q\right\} _{q,\nu}{\rm d}q-{\rm d}H_{1}\right)\wedge{\rm d}\nu=0.
\]
This means that we can find a canonical change of variables iff 
\[
\left\{ P,Q\right\} _{p,\nu}{\rm d}p+\left\{ P,Q\right\} _{q,\nu}{\rm d}q={\rm d}H_{1},
\]
that is, iff the differential form $\left\{ P,Q\right\} _{p,\nu}{\rm d}p+\left\{ P,Q\right\} _{q,\nu}{\rm d}q$
is exact. Following a theorem of Riemann, a form is exact on a simple
connected domain if and only if it is closed. The Poisson brackets
should thus satisfy
\begin{equation}
\frac{\partial}{\partial q}\left\{ P,Q\right\} _{p,\nu}=\frac{\partial}{\partial p}\left\{ P,Q\right\} _{q,\nu}.\label{eq:diff4}
\end{equation}
With some simple manipulations, we can write equality (\ref{eq:diff4})
as 
\[
\frac{\partial}{\partial\nu}\left\{ P,Q\right\} _{p,q}=0,
\]
which is satisfied thanks to the relation $\left\{ P,Q\right\} _{p,q}=1$.

We have just proven the existence of $H_{1}$. The new Hamiltonian
can be expressed in terms of the variables $(P,Q,V',\nu')$ as
\[
H(p,q,V,\nu)=\widetilde{H}(P,\nu')+H_{1}(P,Q,\nu')+V'.
\]
Therefore, the time evolution of the canonical variables is 
\begin{align}
\frac{{\rm d}P}{{\rm d}t} & =-\frac{\partial H_{1}}{\partial Q},\nonumber \\
\frac{{\rm d}Q}{{\rm d}t} & =\frac{\partial\widetilde{H}}{\partial P}+\frac{\partial H_{1}}{\partial P}.\label{eq:time1}
\end{align}
But on the other hand, we know that the time evolution satisfies
\begin{align}
\frac{{\rm d}P}{{\rm d}t} & =\frac{\partial P}{\partial\nu}\dot{\nu},\nonumber \\
\frac{{\rm d}Q}{{\rm d}t} & =\frac{\partial\widetilde{H}}{\partial P}+\frac{\partial Q}{\partial\nu}\dot{\nu}.\label{eq:time2}
\end{align}
The identification of the equalities (\ref{eq:time1}) and (\ref{eq:time2}),
with $\dot{\nu}=1$, gives the desired result.

\end{document}